\documentclass{elsart}
\usepackage{amsmath,amssymb}
\usepackage{graphicx}
\usepackage{subfigure}
\usepackage{hyperref}

\newcommand{\mps}{M_\mathrm{PS}}

\newcommand{\tr}{\mathrm{Tr}}
\newcommand{\re}{\mathrm{Re}}
\newcommand{\im}{\mathrm{Im}}

\begin{document}

\begin{frontmatter}

\title{A Mixed Action Analysis of $\eta$ and $\eta'$ Mesons}

\begin{center}
  \includegraphics[draft=false,width=.2\linewidth]{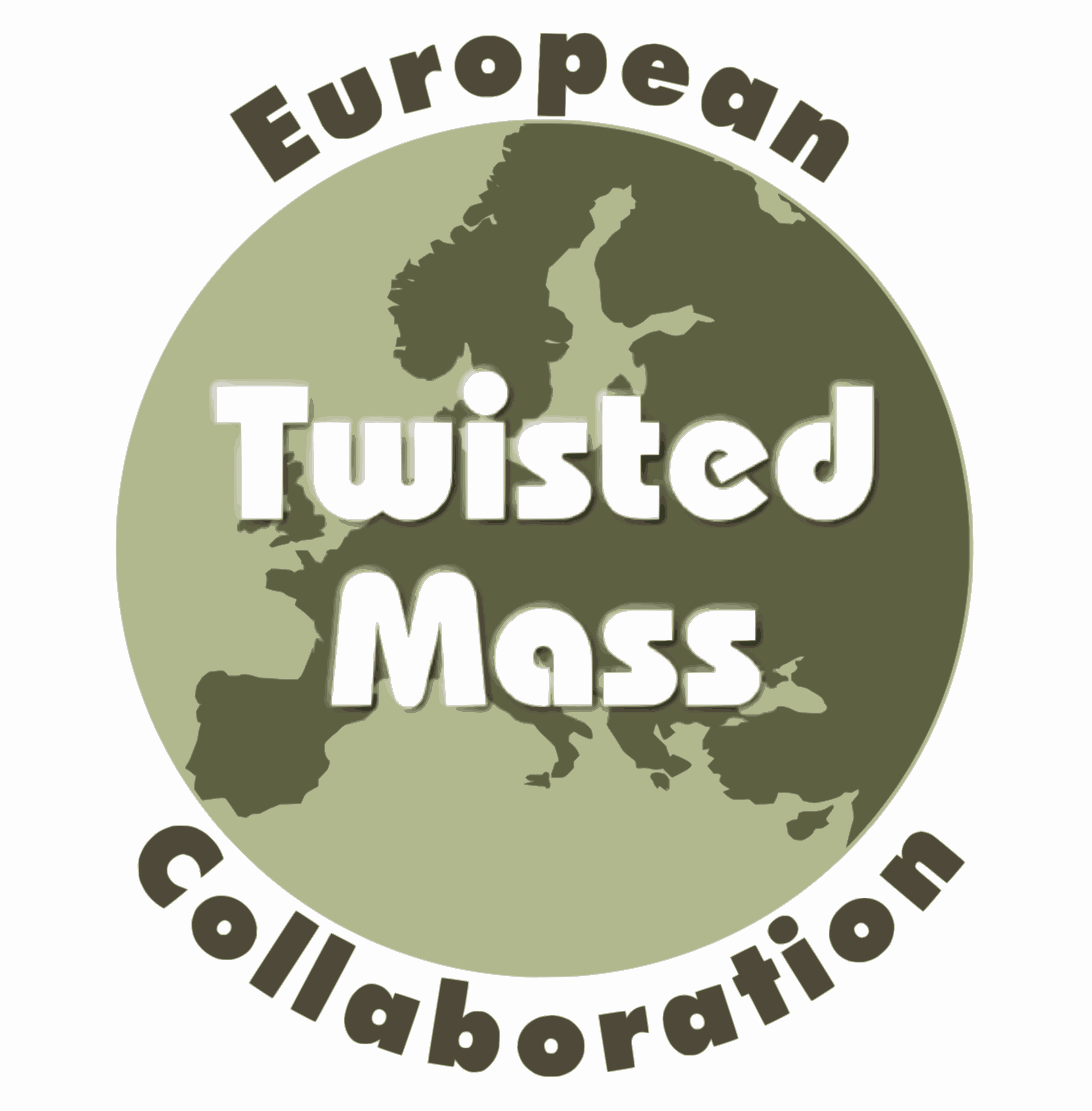}
  \end{center}

  \author{K.~Ottnad, C.~Urbach, F.~Zimmermann} 
  \author{for the ETM Collaboration}

\begin{abstract}
  We study $\eta$ and $\eta'$ mesons and their mixing angle in a mixed
  action approach with so-called Osterwalder-Seiler valence quarks on
  a Wilson twisted mass sea. The gauge configurations have been generated by
  ETMC for $N_f=2+1+1$ dynamical quark flavours and for three values
  of the lattice spacing. The main results are that differences in
  between the mixed action and the unitary approach vanish towards the
  continuum limit with the expected rate of $\mathcal{O}(a^2)$. The
  individual size of the lattice artifacts depends, however, strongly
  on the observable used to match unitary and valence
  actions. Moreover, we show that for the $\eta$ mass valence and
  valence plus sea quark mass dependence differ significantly. Hence, in 
  this case a re-tuning of the simulation parameters in the valence sector 
  only is not sufficient to compensate for mismatches in the original quark masses.
\end{abstract}


\end{frontmatter}

\section{Introduction}

{\it Mixed action} approaches, where valence and sea fermion actions are
chosen differently, are used frequently in lattice QCD. They possess a
number of important advantages compared to the so-called {\it unitary}
case, where valence and sea quark actions are identical. In
particular, it might be possible to use a valence action
obeying more symmetries than the sea action in cases where the valence
action cannot be used in the sea for theoretical reasons or because of
too high computational costs~\cite{Bar:2002nr}. Prominent examples are
overlap~\cite{Neuberger:1997fp,Neuberger:1998wv,Bar:2010ix} 
or domain wall~\cite{Kaplan:1992bt} valence quarks on a Wilson-like or staggered
sea. Concrete examples can be found for instance in
Refs.~\cite{Bowler:2004hs,Cichy:2012vg,Carrasco:2014cwa}.

When working in a mixed action approach, valence and sea actions
need to be matched appropriately, for instance by tuning the valence
quark masses such that a choice of meson masses agrees between
unitary and mixed approaches.
People even go one step further and try to correct for small
mismatches in bare parameters used in the sea
by using a partially quenched mixed action approach. The most extreme
example for this approach is to use valence strange and charm on gauge
configurations with only $N_f=2$ light dynamical quark flavours.
In this case sea and valence actions are not matched, but the valence
parameters are tuned such as to reproduce a choice of physical
observables. 

This ansatz has the big advantage that the gauge configurations do not
need to be re-generated. However, while apparently quite successful,
it is questionable whether this procedure works for observables with a
strong sea quark dependence. Due to OZI suppression there are not many
examples of such observables. But their very existence makes a clear
distinction between QCD and the naive quark model.

Of course, a mixed action approach has also disadvantages, most
prominently the breaking of unitarity, which might for instance drive
certain correlators
negative~\cite{Bardeen:2001jm,Bardeen:2003qz}. Also, it is not clear a
priori how big lattice artifacts one encounters in mixed formulations. 

In this paper we will present results on a particular mixed
action approach with so-called
Osterwalder-Seiler~\cite{Frezzotti:2004wz} valence quarks on an
$N_f=2+1+1$ flavour Wilson twisted mass sea~\cite{Frezzotti:2000nk}.
This particular action combination has the advantage that exact
valence quark flavour symmetry is preserved. Moreover, the respective
zero modes of sea and valence quarks coincide in the chiral
limit. However, $\mathcal{O}(a^2)$ violations of flavour (and parity)
stemming from the sea quarks are still reflected in the magnitude of
lattice artifacts on various physical observables. 

As physical example we study the $\eta$ and $\eta'$ system. The large 
mass splitting observed among $M_{\pi^0}\ll M_\eta\ll M_{\eta'}$ is
thought to be due to the $U_A(1)$ anomaly, a relation established via
the Witten-Veneziano
formula~\cite{Witten:1979vv,Veneziano:1979ec,Veneziano:1980xs}. This
lets one expect a 
significant dependence on the sea quark degrees of freedom. Speaking
more technically, the corresponding correlation functions obtain 
significant contributions from fermionic disconnected diagrams and
are, therefore, uniquely sensitive to differences between valence
and sea formulations. Note that this was also discussed in the context
of the validity of the fourth root trick in staggered
simulations, see Refs.~\cite{Gregory:2006wk,Gregory:2011sg} and
references therein.

After matching valence and sea actions, we compare observables
extracted from unitary and valence operators. The unitary observables
have been computed in Refs.~\cite{Ottnad:2012fv,Michael:2013gka}. 
We study the continuum limit with different matching conditions and
find remarkably good agreement to the unitary case. However, when
comparing the valence with valence plus sea strange quark mass
dependence of $M_\eta$ we find significant differences. 

These findings are important for future lattice QCD investigations:
there are many phenomenologically interesting quantities involving
flavour singlet pseudo-scalar mesons, for instance form factors of $B$
or $D_s$ decays to $\eta\ell\nu$. And maybe most prominently, there are
anomaly related form factors of $\eta\to\gamma\gamma$, which can
be used to estimate the light-by-light contribution to the hadronic
part in the anomalous magnetic moment of the muon, in which we
currently observe a deviation between theory and
experiment at the few $\sigma$
level~\cite{Bennett:2006fi,Aoyama:2012wk,Davier:2010nc,Hagiwara:2011af}.
The usage of Wilson twisted mass fermions described in this paper has
significant advantages compared to other lattice actions due to a
powerful variance reduction. And the possibility to use a mixed action
will further ease those computations. 

More generally, the findings here show that with a mixed action
approach one can deal with fermionic disconnected diagrams, provided
one applies an appropriate matching procedure. These disconnected
diagrams become more and more important as they need to be treated
appropriately for instance in investigations of hadron-hadron
interactions. Since we show here that a mixed action approach works in
the case of $\eta, \eta'$ mesons, where the fermionic disconnected
diagrams contribute significantly, we are confident that the same
approach can be used for other physical observables. First accounts 
of this work can be found in Ref.~\cite{Cichy:2012hq}. Other studies
of $\eta$ and $\eta'$ mesons from lattice QCD can be found in
Refs.~\cite{Gregory:2011sg,Christ:2010dd,Kaneko:2009za,Dudek:2011tt,Dudek:2013yja,Bali:2014pva}.

\section{Lattice actions}
\label{sec:actions}

\begin{table}[t!]
 \centering
 \begin{tabular*}{1.\textwidth}{@{\extracolsep{\fill}}lccccccc}
  \hline\hline
  ensemble & $\beta$ & $a\mu_\ell$ & $a\mu_\sigma$ & $a\mu_\delta$ &
  $V$ & $N_\mathrm{conf}$  & $N_b$ \\ 
  \hline\hline
  $A40.24$   & $1.90$ & $0.0040$ & $0.150$  & $0.190$  &
  $24^3\times48$ & $1117$ & $5$ \\
  $A60.24$   & $1.90$ & $0.0060$ & $0.150$  & $0.190$  &
  $24^3\times48$ & $1249$ & $5$ \\
  $A80.24$   & $1.90$ & $0.0080$ & $0.150$  & $0.190$  &
  $24^3\times48$ & $2441$ & $10$ \\
  $A100.24$  & $1.90$ & $0.0100$ & $0.150$  & $0.190$  &
  $24^3\times48$ & $968$ & $5$ \\
  \hline
  $A80.24s$  & $1.90$ & $0.0080$ & $0.150$  & $0.197$  &
  $24^3\times48$ & $2420$ & $10$ \\
  $A100.24s$ & $1.90$ & $0.0100$ & $0.150$  & $0.197$  &
  $24^3\times48$ & $1196$ & $5$ \\
  \hline
  $B55.32$   & $1.95$ & $0.0055$ & $0.135$  & $0.170$  &
  $32^3\times64$ & $4450$ & $5$ \\
  $D45.32sc$ & $2.10$ & $0.0045$ & $0.0937$ & $0.1077$ &
  $32^3\times64$ & $2220$ & $10$ \\
  \hline\hline
 \end{tabular*}
 \caption{The gauge ensembles used in this study. For the labelling of
   the ensembles we adopted the notation in
   Ref.~\cite{Baron:2010bv}. In addition to the relevant input
   parameters we give the lattice volume, the number of evaluated
   configurations $N_\mathrm{conf}$ and the block length $N_b$ used
   for bootstrapping. $N_b$ was chosen such that blocks cover at least
   20 HMC trajectories of length one.}   
 \label{tab:setup}
\end{table}

The results we will present here are obtained by
evaluating correlation functions on gauge configurations provided by the European Twisted Mass 
Collaboration (ETMC)~\cite{Baron:2010bv}. We use the ensembles
specified in table~\ref{tab:setup} adopting the notation from
Ref.~\cite{Baron:2010bv}. More details can be found in this
reference. 

The sea quark formulation is the Wilson twisted mass formulation with
$N_f=2+1+1$ dynamical quark flavours. The Dirac operator for the light
quark doublet reads~\cite{Frezzotti:2000nk} 
\begin{equation}
  D_\ell = D_W + m_0 + i \mu_\ell \gamma_5\tau^3\, ,
  \label{eq:Dlight}
\end{equation}
where $D_W$  denotes the standard Wilson Dirac operator and $\mu_\ell$
the bare light twisted mass parameter. $\tau^3$ and in general
$\tau^i, i=1,2,3$ represent the Pauli matrices acting in flavour
space. $D_\ell$ acts on a spinor $\chi_\ell = (u,d)^T$ and, hence, the $u$
($d$) quark has twisted mass $+\mu_\ell$ ($-\mu_\ell$).

For the heavy unitary doublet of $c$ and
$s$ quarks~\cite{Frezzotti:2003xj} the Dirac operator is given by
\begin{equation}
  D_h = D_W + m_0 + i \mu_\sigma \gamma_5\tau^1 + \mu_\delta \tau^3\,.
  \label{eq:Dsc}
\end{equation}
The bare Wilson quark mass $m_0$ has been tuned to its critical
value~\cite{Chiarappa:2006ae,Baron:2010bv}. This guarantees
automatic order $\mathcal{O}\left(a\right)$ improvement
\cite{Frezzotti:2003ni}, which is one of the main advantages of the
Wilson twisted mass formulation of lattice QCD. 

$\eta$ and $\eta'$ masses have been computed in this framework in
Refs.~\cite{Ottnad:2012fv,Michael:2013gka,OttnadPhd2014} on the same
set of gauge configurations used here (and more) -- we will refer to
this framework as the \emph{unitary approach}. However, in order to
account for -- and possibly benefit from -- correlations we have
re-evaluated the unitary $\eta$ and $\eta'$ masses on exactly the same
gauge configurations as used in the present study.

The splitting term in the heavy doublet (\ref{eq:Dsc}) introduces flavour mixing
between strange and charm quarks which needs to be accounted for in
the analysis. However, this complication can be avoided by using a
mixed action approach for the valence strange and charm quarks. 
Formally, we introduce so-called Osterwalder-Seiler (OS) twisted
valence strange and charm
quarks~\cite{Frezzotti:2004wz,Blossier:2007vv}. The Dirac operator  
for a single valence quark flavour $q$ reads
\begin{equation}
  \label{eq:Dos}
  D_q = D_W + m_0 + i \mu_q \gamma_5\,.
\end{equation}
Adapting the ideas of Ref.~\cite{Frezzotti:2004wz} to the $\eta,\eta'$
system, we introduce two strange and two charm quark flavours, $s$,
$s'$ and $c$, $c'$, respectively. Flavours $s$ and $s'$ will have
quark mass with equal modulus, but opposite sign: $\mu_s = |\mu_s| = -\mu_{s'}$,
and the same for $c$ and $c'$. Formally, the lattice action is extended to include
a fermionic action corresponding to the Dirac operators
(\ref{eq:Dos}) for all valence strange and charm quark flavours,
accompanied by a ghost action to exactly cancel the contributions of
the additional valence quarks to the fermionic determinant. For
details we refer to Ref.~\cite{Frezzotti:2004wz}. In this reference it
was also shown that automatic $\mathcal{O}(a)$-improvement stays valid
in this framework and unitarity is restored in the continuum limit. In
particular, flavours $s$ and $s'$ ($c$ and $c'$) become identical.

It is important to notice that at finite lattice spacing
values correlation functions involving $s$ and $s'$ ($\mu_s =
-\mu_{s'}$) differ by lattice artifacts. For instance, the masses
extracted from the correlation function of the operator 
\begin{equation}
  \label{eq:OK+}
  \mathcal{O}_{K}^\mathrm{OS} =\bar\psi_s i\gamma_5 \psi_d
\end{equation}
where the fields $\psi_q$, $\bar{\psi}_q$ denote single quark fields in the so-called physical basis, differ from the one extracted from the operator
\begin{equation}
  \label{eq:OKOS}
  \mathcal{O}_{K^0}^\mathrm{OS} = \bar\psi_{s'} i\gamma_5 \psi_d
\end{equation}
by $\mathcal{O}(a^2)$ (we denote it with $K^0$ in remedy of the
neutral pion in the light sector). Only in the continuum limit these two masses
will agree again.

\begin{table}[t!]
 \centering
 \begin{tabular*}{.8\textwidth}{@{\extracolsep{\fill}}lccc}
  \hline\hline
  ensemble & $aM_{K}$ & $aM_{\eta_s}$ & $aM_{\pi^0_\mathrm{conn}}$  \\
  \hline\hline
  $A40.24$   & $0.25884(43)$ & $0.30708(60)$ & $0.2375(25)$ \\
  $A60.24$   & $0.26695(52)$ & $0.31010(65)$ & $0.2544(26)$ \\
  $A80.24$   & $0.27706(61)$ & $0.31406(46)$ & $0.2659(25)$ \\
  $A100.24$  & $0.28807(34)$ & $0.31575(45)$ & $0.2883(14)$ \\
  \hline
  $A80.24s$  & $0.25503(33)$ & $0.27168(49)$ & $0.2649(16)$ \\
  $A100.24s$ & $0.26490(74)$ & $0.27455(73)$ & $0.2841(16)$ \\
  \hline
  $B55.32$   & $0.22799(34)$ & $0.26087(33)$ & $0.2177(10)$ \\
  $D45.32sc$   & $0.17570(84)$ & $0.21126(34)$ & $0.1494(15)$ \\
  \hline\hline
 \end{tabular*}
 \caption{Values of the unitary kaon and $\eta_s$ masses $M_K$ and $M_{\eta_s}$ in lattice units, which have
   been used to match mixed and unitary actions. In addition, we give the mass values $M_{\pi^0_\mathrm{conn}}$ of the connected
   neutral pion which becomes identical to $\eta_s$ for mass degenerate light and strange quarks. This SU$(3)$
   symmetric situation is realised approximately for the $A80.24s$ and
   $A100.24s$ ensembles. The kaon mass data shown in this table has first been published in Ref.~\cite{Ottnad:2012fv}.}
 \label{tab:matchedUSObs}
\end{table}

Valence and unitary actions need to be matched appropriately. 
As shown in Ref.~\cite{Frezzotti:2004wz}, in our case
the matching can be performed in principle using the relation
\begin{equation}
  \label{eq:musc}
  \mu_{c/s} = \mu_\sigma\ \pm\ Z_P/Z_S\ \mu_\delta \,.
\end{equation}
However, for the strange quark mass uncertainties in $Z_P/Z_S$ are
magnified in $a\mu_s$, and thus we decided not to rely on
Eq.~\ref{eq:musc}. Instead, meson masses are used:
in previous studies it was found that matching kaon
masses determined from the operator (\ref{eq:OK+}) to the unitary kaon
masses is best in the sense that the residual lattice artifacts in the
results computed in a mixed action approach are
small~\cite{Frezzotti:2005gi}. We will call this procedure kaon
matching. 

For details on how to compute the kaon mass in the unitary case we
refer to Ref.~\cite{Baron:2010th}. We note in passing that there is
no kaon mass splitting introduced by the twisted mass formalism in the
unitary case for the choice of a degenerate light quark doublet $|\mu_u|=|\mu_d|=\mu_\ell$~\cite{Chiarappa:2006ae}.

As a second matching observable for the strange quark mass we use the
mass of the so-called $\eta_s$ meson $M_{\eta_s}$. The $\eta_s$ is an artificial meson corresponding to the following interpolating operator 
\begin{equation}
 \mathcal{O}_{\eta_s}^\mathrm{OS} = \bar{\psi}_s i\gamma_5 \psi_{s''}\,,
\end{equation}
for which we assume $\mu_s=\mu_{s''}$, unlike the $s'$ quark considered above, which had opposite sign. A benefit of this 
particular choice is the absence of disconnected diagrams in the corresponding 
two-point function. This procedure will be called $\eta_s$ matching. For
technical details, e.g. further interpolating fields and correlation functions
we refer to section~\ref{sec:mesons}. 

For both matching procedures on each gauge ensemble one tunes the value of $a\mu_s$ such that the kaon or the 
$\eta_s$ mass agrees within errors between the mixed and the
unitary formulation. The unitary values of the masses we matched to
are compiled in table~\ref{tab:matchedUSObs}.
In order to compute the matching values
for $a\mu_s$ we performed inversions on a subset of the available configurations in a range of $a\mu_s$ values 
and interpolated the squared OS meson masses linearly in $a\mu_s$. The
matching values for $a\mu_s$ for the two matching observables and all
ensembles can be found in table~\ref{tab:quarkmasses}. The values for
$M_K^\mathrm{OS}$ and $M_{\eta_s}^\mathrm{OS}$ at the matching points
are compiled in the appendix in table~\ref{tab:OSmasses}. Note that in 
case of matching $M_{\eta_s}$ we do not reach exact agreement for all ensembles 
within errors when recomputing $M_{\eta_s}^\mathrm{OS}$ from full statistics. 
These numerically small differences become irrelevant for the $\eta$ 
and $\eta'$ masses themselves due to the much larger statistical
uncertainties introduced by the quark disconnected diagrams .

In the following we indicate quantities determined in the OS framework
with the superscript $^\mathrm{OS}$, while quantities determined in
the unitary case have no superscript. To distinguish the two matching
procedures we use the superscripts $^K$ and $^{\eta_s}$.

\begin{figure}[t]
  \centering
  \includegraphics[width=0.6\linewidth]{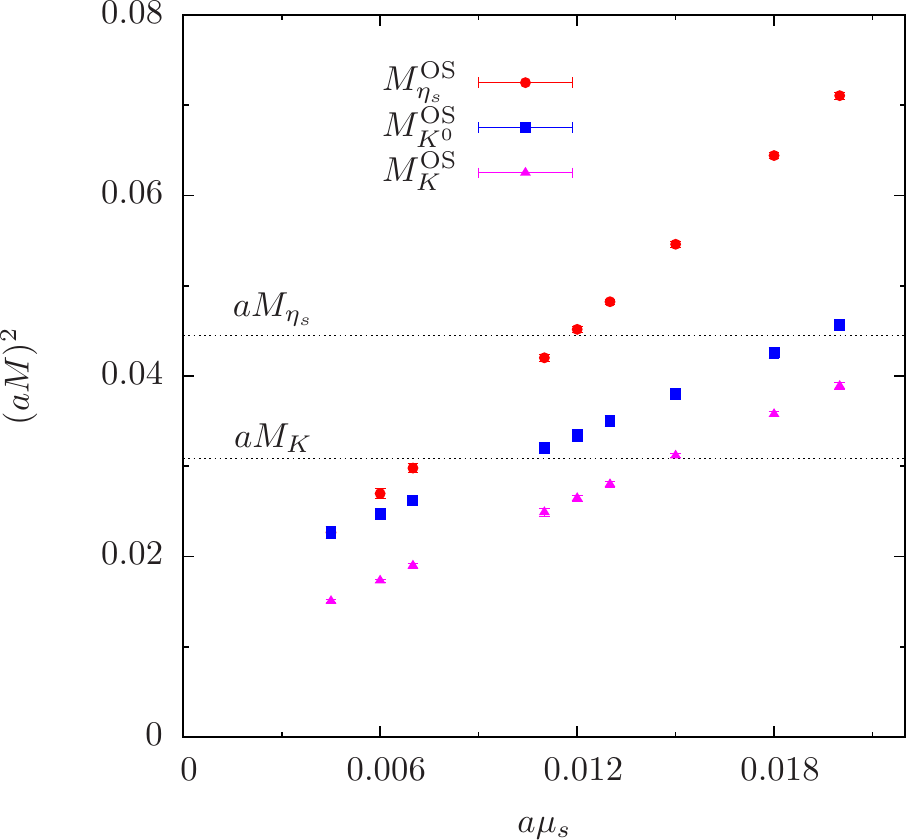}
  \caption{$(aM_{K}^\mathrm{OS})^2$, $(aM_{\eta_s}^\mathrm{OS})^2$ and
    $(aM_{K^0}^\mathrm{OS})^2$ as a function of the 
    bare OS strange quark mass $a\mu_s$ for the $D45.32sc$
    ensemble. Horizontal lines indicate the unitary mass values that
    have been used for the matching.}
  \label{fig:D45matching}
\end{figure}

As an example for the matching procedure we show in
figure~\ref{fig:D45matching} $(aM_{K}^\mathrm{OS})^2$, 
$(aM_{\eta_s}^\mathrm{OS})^2$ and $(aM_{K^0}^\mathrm{OS})^2$ as a function of the
bare OS strange quark mass $a\mu_s$ for the $D45.32sc$ ensemble. In
figure~\ref{fig:D45matching} the aforementioned OS kaon mass splitting
can be observed. In the limit $\mu_s = \mu_\ell$ this splitting
corresponds to the difference between the charged pion mass and the
connected only neutral pion mass. The splitting is almost independent
of $\mu_s$, decreasing slightly with increasing $\mu_s$. 

As expected, $M_{\eta_s}$ is larger than the two kaon masses and agrees
with $M_{K^0}^\mathrm{OS}$ in the limit $\mu_s=\mu_\ell$. All three
squared masses show a linear dependence on $\mu_s$. The horizontal
lines indicate the corresponding unitary values that have been used
for computing $a\mu_s^{K}$ and $a\mu_s^{\eta_s}$. 

For the charm quark mass the estimate from Eq.~\ref{eq:musc} is less
affected by uncertainties. In order to circumvent the need for
$Z_P/Z_s$, one can re-arrange Eq.~\ref{eq:musc} to
\begin{equation}
\mu_c = 2\mu_\sigma - \mu_s\,. \label{eq:quark_masses}
\end{equation}
Because $\mu_s \ll \mu_c$ and $\eta, \eta'$ do not depend on
$\mu_c$, we restrict ourselves to kaon matching for the $\mu_s$ value entering the charm
quark mass. The corresponding values for $\mu_c\equiv\mu^{K}_c$
extracted in this way can be found in table~\ref{tab:quarkmasses}.

\begin{table}[t!]
 \centering
 \begin{tabular*}{.9\textwidth}{@{\extracolsep{\fill}}lc|ccc}
  \hline\hline
  ensemble & $a\mu_\ell$ & $a\mu^{K}_s$ &  $a\mu^{\eta_s}_s$ & $a\mu^{K}_c$ \\ 
  \hline\hline
  $A40.24$   & $0.0040$ & $0.02300(25)$ & $0.01239(25)$ & $0.27700(25)$ \\
  $A60.24$   & $0.0060$ & $0.02322(22)$ & $0.01303(22)$ & $0.27678(22)$ \\
  $A80.24$   & $0.0080$ & $0.02328(20)$ & $0.01338(20)$ & $0.27672(20)$ \\
  $A100.24$  & $0.0100$ & $0.02381(21)$ & $0.01380(22)$ & $0.27619(21)$ \\
  \hline
  $A80.24s$  & $0.0080$ & $0.01884(16)$ & $0.00883(21)$ & $0.28116(16)$ \\
  $A100.24s$ & $0.0100$ & $0.01877(22)$ & $0.00922(23)$ & $0.28123(22)$ \\
  \hline
  $B55.32$   & $0.0055$ & $0.01858(12)$ & $0.01100(10)$ & $0.25142(12)$ \\
  $D45.32sc$ & $0.0045$ & $0.01488(30)$ & $0.01180(12)$ & $0.17252(30)$ \\
  \hline\hline
 \end{tabular*}
 \caption{Matching values of the OS valence strange quark masses
   $\mu_s$ for kaon and $M_{\eta_s}$ matching. The OS valence charm
   quark masses $\mu^{K}_c$ have been determined using Eq.~(\ref{eq:quark_masses}) for kaon matching
   only.} 
 \label{tab:quarkmasses}                                                                                                          
\end{table}

All errors are computed using a blocked bootstrap procedure
to account for autocorrelation as well as all other statistical correlations in the data. The number of bootstrap samples was
taken to be $1000$ and the number of configurations per block $N_b$ is
given for every ensemble in table~\ref{tab:setup}. $N_b$ itself was
chosen such that the length of a block corresponds to at least $20$
HMC trajectories of length one. This value turned out sufficient to
compensate for autocorrelation in the observables considered in this
study.

\section{Pseudo-scalar flavour-singlet mesons} \label{sec:mesons}

In order to extract $\eta$ and $\eta'$ states we need a set of
appropriate interpolating operators. As we are going
to work in the quark flavour basis our choice is
\[
\begin{split}
  \mathcal{O}_\ell^p(t) &= \frac{1}{\sqrt{2}} \sum_\mathbf{x}(\bar
  \psi_u i\gamma_5 \psi_u(\mathbf{x},t) + \bar \psi_d
i\gamma_5 \psi_d(\mathbf{x},t))\,,\\
  \mathcal{O}_s^p(t) &=  \sum_\mathbf{x}\bar \psi_s\,
  i\gamma_5\, \psi_s(\mathbf{x},t)\,,\\
  \mathcal{O}_c^p(t) &= \sum_\mathbf{x}\bar \psi_c\,
  i\gamma_5\, \psi_c(\mathbf{x},t)
\end{split}
\]
in the physical basis, again denoted as $\bar\psi_q,\psi_q$. With
Osterwalder-Seiler valence 
fermions we have to rotate the bilinears into the so-called twisted
basis denoted as $\bar q,q$, see e.g. Ref.~\cite{Shindler:2007vp}, in
which also the Dirac 
operators in the previous section were written. Performing this axial
rotation~\cite{Frezzotti:2000nk,Frezzotti:2004wz}, one obtains the
following operators in the so-called twisted basis
\[
\begin{split}
  \mathcal{O}_\ell(t) &= \frac{1}{\sqrt{2}} \sum_\mathbf{x}(\bar{d} d(\mathbf{x},t) - \bar{u} u(\mathbf{x},t))\,,\\
  \mathcal{O}_s(t) &=  -\sum_\mathbf{x}\bar s
  s(\mathbf{x},t)\,,\\
  \mathcal{O}_c(t) &= -\sum_\mathbf{x}\bar c
  c(\mathbf{x},t)\,.
\end{split}
\]
From these operators we build a correlation function matrix
\begin{equation}
 \label{eq:correlations}
 \mathcal{C}(t)_{qq'} =
 \langle\mathcal{O}_q(t'+t)\ \mathcal{O}_{q'}^\dagger(t')\rangle\,,\quad
 q,q'\in\{\ell,s,c\}\,,
\end{equation}
which allows us to obtain results for masses and amplitudes; cf. section~\ref{subsec:matrix_of_correlation_functions}.

The corresponding correlation functions have fermionic connected and
disconnected contributions. The case for up and down quarks is like in
the unitary approach and discussed in detail in
Refs.~\cite{Jansen:2008wv,Ottnad:2012fv,Michael:2013gka}. Therefore,
we concentrate on the disconnected contributions for strange and charm
quarks. The correlation function of
$\mathcal{O}_s(t)$, for instance, has the following contributions
\begin{equation}
  \label{eq:Oscor}
  \langle  \mathcal{O}_s(t)\, \mathcal{O}_s^\dagger(0) \rangle_F = - \tr\{G_s^{0t}
    G_s^{t0}\} + \tr\{G_s^{tt}\}\cdot\tr\{G_s^{00}\}\,,
\end{equation}
where $\langle . \rangle_F$ denotes the average of fermions only and
\begin{equation}
  \label{eq:Gs}
  G_s^{xy}=(D_s^{-1})(x,y)
\end{equation}
denotes the strange OS propagator. The first term in Eq.~\ref{eq:Oscor} is
the connected contribution and the second the disconnected one. Note
that mixed flavour correlation functions have only disconnected
contributions by definition. The ground state mass extracted only from the 
connected piece on the r.h.s. of Eq.~\ref{eq:Oscor} is the mass of the 
artificial $\eta_s$ meson, which is employed for $\eta_s$ matching.

We evaluate the connected only contribution to Eq.~\ref{eq:Oscor}
using the one-end-trick~\cite{Boucaud:2008xu}. In contrast
to the Wilson case, $\tr\{G_s^{0t} G_s^{t0}\}$ is in general complex
valued. However, the imaginary part of the corresponding trace is a
pure lattice artifact. This can be shown by considering a suitable
combination of connected correlation functions involving OS quarks $s$
and $s'$ 
\[
\begin{split}
\langle \bar\psi_s i\gamma_5 \psi_s(x)\bar\psi_s i\gamma_5 \psi_s(0) & - \{s\rightarrow s'\} \rangle_F = \langle \bar{s}s(x)\bar{s}s(0) - \bar{s'}s'(x)\bar{s'}s'(0) \rangle_F \\
=& -\tr\{G_{s}^{0x}G_{s}^{x0}\} + \tr\{G_{s'}^{0x}G_{s'}^{x0}\} \\
=& -\tr\{G_{s}^{0x}\gamma_5 (G_{s'}^{0x})^\dag \gamma_5 \} + \tr\{G_{s'}^{0x}\gamma_5 (G_{s}^{0x})^\dag\gamma_5 \} \\
=& -\tr\{G_{s}^{0x}\gamma_5 (G_{s'}^{0x})^\dag \gamma_5 \} + \tr\{G_{s}^{0x}\gamma_5 (G_{s'}^{0x})^\dag\gamma_5 \}^\dag \\
=& -2i\,\im\tr\{G_{s}^{0x}\gamma_5 (G_{s'}^{0x})^\dag \gamma_5 \} \\
=& -2i\,\im\tr\{G_{s}^{0x}G_{s}^{x0} \} \,,
\end{split}
\]
where we have used the relation $D_s = \gamma_5D_{s'}^\dagger
\gamma_5$ together with the cyclic property of the trace. Since the
l.h.s. of the above relation vanishes in the continuum limit, we will
drop the imaginary part in our calculations. 

For the disconnected contribution to Eq.~\ref{eq:Oscor} we need to
estimate $\tr\{G_s^{tt}\}$ on every gauge configuration and all
$t$-values. $\tr\{G_s^{tt}\}$ is again in general complex valued. And
again, one can show that the real part is a pure lattice
artifact. Similar to the case of the connected contribution this can
be inferred from the following equality
\[
\begin{split}
  \langle -\bar\psi_s i\gamma_5 \psi_s(x) +\bar\psi_{s'} i\gamma_5
  \psi_{s'}(x)\rangle_F &= \langle \bar{s} s(x) + \bar{s}' s'(x)\rangle_F \\
  &= -\tr\{G_s^{xx}\} - \tr\{G_{s'}^{xx}\} \\
  &= -\tr\{G_s^{xx}\} - \tr\{G_s^{xx}\}^\dagger \\
  &=-2\,\re\,\tr\{G_s^{xx}\}\,,
\end{split}
\]
which is zero in the continuum limit. Therefore, we will also
drop the real part of the disconnected loops in the
calculation. Similarly one can show that
\begin{equation}
  \label{eq:OsIm}
  \langle \bar{s} s(x) - \bar{s}' s'(x)\rangle_F
  =-2i\,\im\,\tr\{G_s^{xx}\}\,.
\end{equation}
We remark that all of the above results hold for any further valence
quark as well (e.g. the charm quark).

The full strange correlation function after subtraction of lattice
artifacts is then given as
\begin{equation}
  \langle  \mathcal{O}_s(t)\, \mathcal{O}_s^\dagger(0) \rangle_F = - \re\tr\{G_s^{0t}
    G_s^{t0}\} - \im\,\tr\{G_s^{tt}\}\cdot\im\,\tr\{G_s^{00}\}\,,
\end{equation}
and analogously for the charm. Cross flavour terms involve only disconnected
diagrams and are for instance given as
\begin{equation}
  \langle\mathcal{O}_s(t)\, \mathcal{O}_c^\dagger(0) \rangle_F = -
  \im\,\tr\{G_s^{tt}\}\cdot\im\tr\{G_c^{00}\}\,.
\end{equation}

\subsection{Variance Reduction}
\label{sec:variance}

The relation (\ref{eq:OsIm}) enables us to use a very powerful
variance reduction method developed originally for the disconnected
contributions of the light doublet~\cite{Jansen:2008wv} also for
strange and charm flavours (see also
Ref.~\cite{Alexandrou:2013wca}). It is based on 
the identity (recall $\mu_s = -\mu_{s'}$) 
\[
D_s^{-1} - D_{s'}^{-1} = -2i\mu_s D_{s'}^{-1}\ \gamma_5\ D_s^{-1}\,.
\]
Therefore, using Eq.~\ref{eq:OsIm} we can estimate 
\begin{equation}
  \label{eq:oneend}
  \im\,\tr\{G_s^{xx}\} = -\mu_s
  \tr\{G_{s'}^{xy}\ \gamma_5\ G_s^{yx}\} \,,
\end{equation}
and correspondingly for the charm quark. Following
Ref.~\cite{Jansen:2008wv}, we apply this variance reduction method
also to the light doublet. 

\subsection{Matrix of Correlation Functions}
\label{subsec:matrix_of_correlation_functions}
By applying these results, we compute the matrix of Euclidean correlation
functions in Eq.~(\ref{eq:correlations}) and solve the generalised eigenvalue problem (GEVP) \cite{Michael:1982gb,Luscher:1990ck,Blossier:2009kd} 
\begin{equation}
  \mathcal{C}(t)\ v^{(n)}(t,t_0) = \lambda^{(n)}(t, t_0)\
  \mathcal{C}(t_0)\ v^{(n)}(t,t_0) \,,
  \label{eq:gevp}
\end{equation}
for determining the meson masses $M_\eta$, $M_{\eta'}$ (and possibly
$M_{\eta_c}$) from the principal correlators
$\lambda^{(n)}(t,t_0)$, $n\in \eta,\eta'$. The effective masses are then
computed by numerically solving
\[
\frac{ \lambda^{(n)}(t,t_0) } { \lambda^{(n)}(t+1,t_0) }
=
\frac{ \exp^{-M_\mathrm{eff}^{(n)}t}+\exp^{-M_\mathrm{eff}^{(n)}(T-t)} } { \exp^{-M_\mathrm{eff}^{(n)}(t+1)}+\exp^{-M_\mathrm{eff}^{(n)}(T-(t+1))} }
\]
for $aM_\mathrm{eff}^{(n)}$. The matrix $\mathcal{C}$ is enlarged to a
$6\times6$ matrix by using in addition fuzzed~\cite{Lacock:1994qx,Boucaud:2008xu} operators.

At this point we recall that in the unitary case also the mass of the $\eta_s$ is not
obtained from a single correlation function but rather from the ground
state of a correlation function matrix involving connected correlation
functions for strange and charm quarks. This minor complication arises
due to the violation of flavour symmetry in the Wilson twisted mass
formulation and the fact that the action can no longer be chosen
flavour-diagonal for a non-degenerate doublet. Therefore, one has to
consider off-diagonal connected correlation functions in addition to
the ones consisting only of strange and charm quarks. However, the
off-diagonal connected contributions are a pure lattice artifact and in the continuum
limit the expected behaviour is restored, i.e. strange and charm sector
decouple regarding the connected pieces.

Apart from meson masses also matrix elements can be extracted from the
GEVP, which are needed to obtain $\eta$ and $\eta'$ mixing angles. 
We define the mixing angles $\phi_\ell$, $\phi_s$ in the quark
flavour basis using the pseudoscalar density matrix elements $A_{q,n} =
\langle 0 | \mathcal{O}_q|n\rangle$ with $n \in
\{\eta ,\ \eta'\}$ and $q \in \{l,\ s\}$ as
\begin{equation}
  \begin{pmatrix}
    A_{\ell, \eta} & A_{s, \eta}\\
    A_{\ell, \eta'} & A_{s, \eta'}\\
  \end{pmatrix}
  =
  \begin{pmatrix}
    c_\ell \cos\phi_\ell & -c_s \sin\phi_s \\
    c_\ell \sin\phi_\ell &  c_s \cos\phi_s \\
  \end{pmatrix}\,,
\end{equation}
see also Refs.~\cite{Ottnad:2012fv,Michael:2013gka}. From chiral
perturbation theory combined with large $N_C$ arguments
$|\phi_\ell-\phi_s|/|\phi_\ell+\phi_s|\ll1$ can be inferred according to
Refs.~\cite{Kaiser:1998ds,Kaiser:2000gs,Feldmann:1998sh,Feldmann:1998vh}
which is confirmed by lattice QCD~\cite{Michael:2013gka}.
Therefore, we will consider only the average mixing angle $\phi$
\begin{equation}
  \tan^2\phi \equiv -\frac{A_{l,\eta'}A_{s,\eta}}{A_{l,\eta}A_{s,\eta'}} \,.
\end{equation}

\subsection{Excited State Removal}

To improve the $\eta'$ (and $\eta$) mass determinations, we use a
method first proposed in Ref.~\cite{Neff:2001zr}, successfully
applied for the $\eta_2$ (the $\eta'$ in $N_f=2$ flavour QCD) in
Ref.~\cite{Jansen:2008wv} and very recently to the $N_f=2+1+1$ case in
Ref.~\cite{Michael:2013gka}. It relies on the assumption that
disconnected contributions are significant only for the $\eta$ and
$\eta'$ state, but negligible for higher excited states. This means,
in turn, that only the connected contributions to $\mathcal{C}$ are
affected by excited states. 

Since the signal-to-noise ratio of the connected contributions is much larger
than the one of the disconnected ones, we can determine the
corresponding ground state at large Euclidean times and subtract the
excited states at small times. This subtracted connected and the full
disconnected contributions are combined in 
$\mathcal{C}^\mathrm{sub}_{q,q'}(t)$, which is then used in the
analysis. We refer to the discussion in Ref.~\cite{Michael:2013gka}
for more details.  

This procedure clearly depends on the validity of the
assumption. However, it can be checked in our Monte-Carlo data: if the
subtracted combination of connected and disconnected contributions
does not show excited states anymore, we take it as a strong hint for
the validity of the assumption. This is the case for all our
ensembles, and it was also the case in the unitary
approach~\cite{Michael:2013gka}.

For all ensembles we observe that the mass of the $\eta$ meson is
unaffected by this procedure within errors. Only the error
estimates get significantly smaller. This is also the case for the
$\eta'$ where, however, the errors are quite significant before
excited state removal.

\begin{figure}[t]
  \centering
  \subfigure[]{\includegraphics[width=0.45\linewidth]{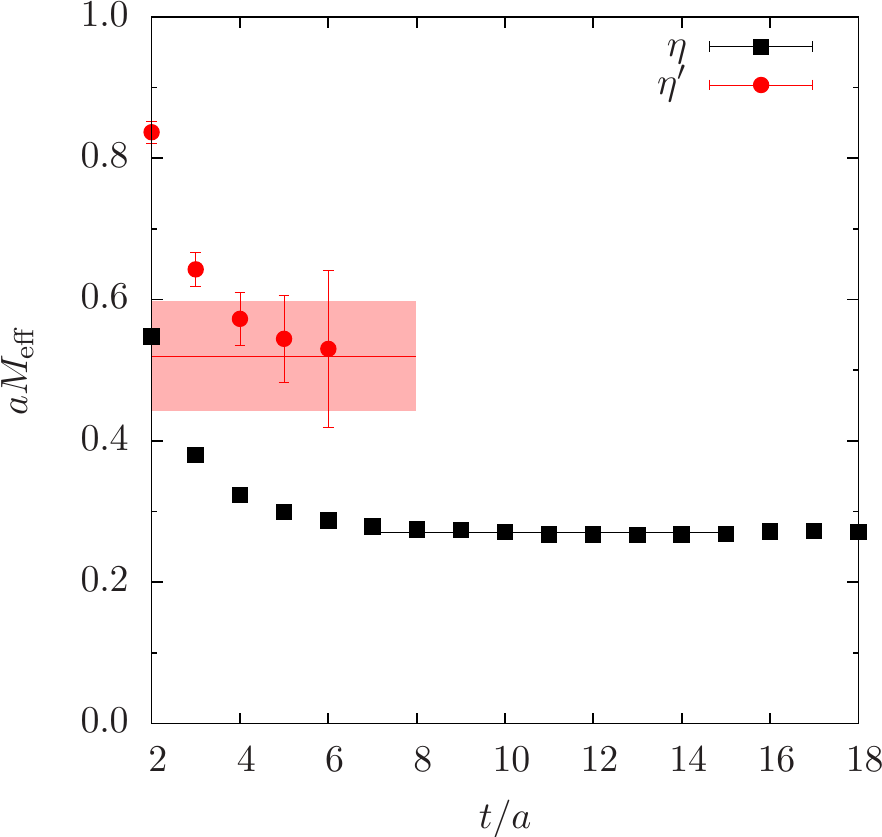}}\quad
  \subfigure[]{\includegraphics[width=0.45\linewidth]{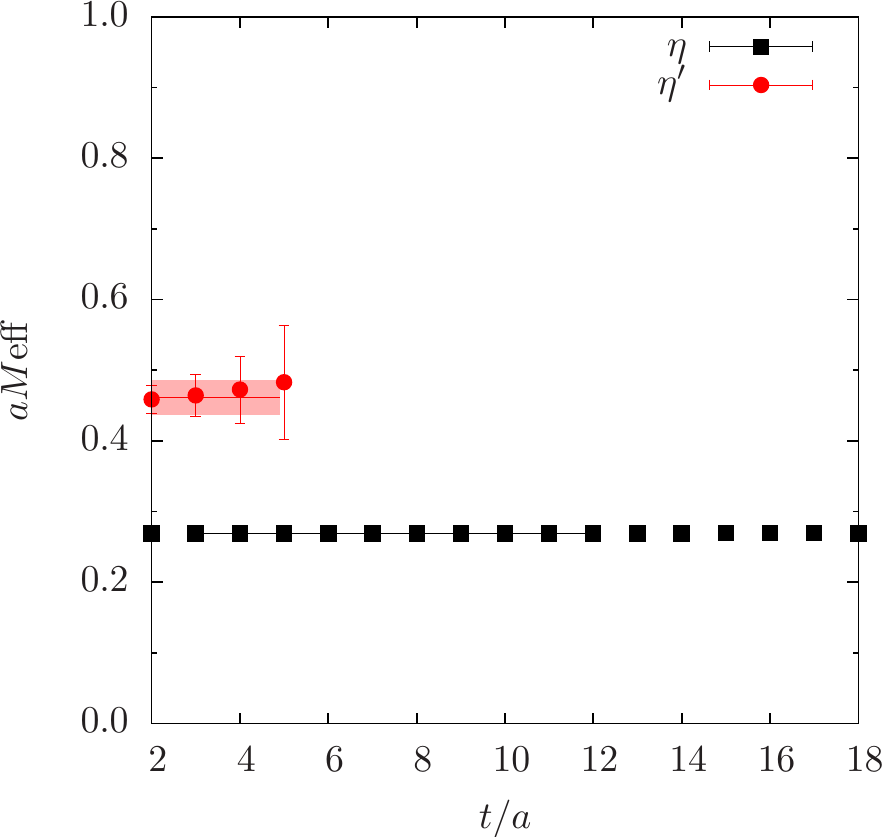}}
  \caption{Effective masses $M_\eta^\mathrm{OS}$ and $M_{\eta'}^\mathrm{OS}$ for the
    $A80.24s$ ensemble using $\eta_s$ matching for (a) a $6\times 6$
    correlation matrix including local and fuzzed operators and (b) a
    $3\times3$ local-only correlation matrix with connected excited
    states subtracted. The fitted values are shown as lines with error
    band. The corresponding fit range is indicated by the length of
    the lines. For further details see text and table~\ref{tab:fits}.} 
  \label{fig:mA80ConnExcTrick}
\end{figure}

As an example we show in figure~\ref{fig:mA80ConnExcTrick} the
effective masses of the two lowest-lying states for the $A80.24s$
ensemble from $\eta_s$ matching. The values shown in the left panel
are obtained from the standard method using a $6\times 6$ correlation
function matrix build from local and fuzzed operators, whereas the
right panel shows the results for a $3\times 3$ local-only correlation
function matrix with excited states removed from its connected
contributions. Our fitted values are always obtained from a
$\cosh$-type fit to the respective principal correlators. The
corresponding fit ranges are indicated by the bands in the plots. In
general, our choice for the fit ranges $\left[t_1,t_2\right]$ and the
values of $t_0$ for the GEVP are given in table~\ref{tab:fits}. We
remark that -- since there is usually no clear plateau reached for the
$\eta'$ state from the standard method -- we apply a two-state fit to
the corresponding principal correlator in this case. In all other
cases we employed a single state fit in the plateau region. 

Comparing the two panels in figure~\ref{fig:mA80ConnExcTrick} one
observes that the $\eta$ mass plateau is unaffected, but starts at
$t/a=2$ in the case of removed excited states. For the $\eta'$ there
is no plateau reached in the left panel before the signal is lost in
noise, whereas in the right one a reasonable plateau is visible. The
extracted masses still agree within errors.

\section{Results}

\begin{figure}[t]
  \centering
  \subfigure[]{\includegraphics[width=.48\linewidth]{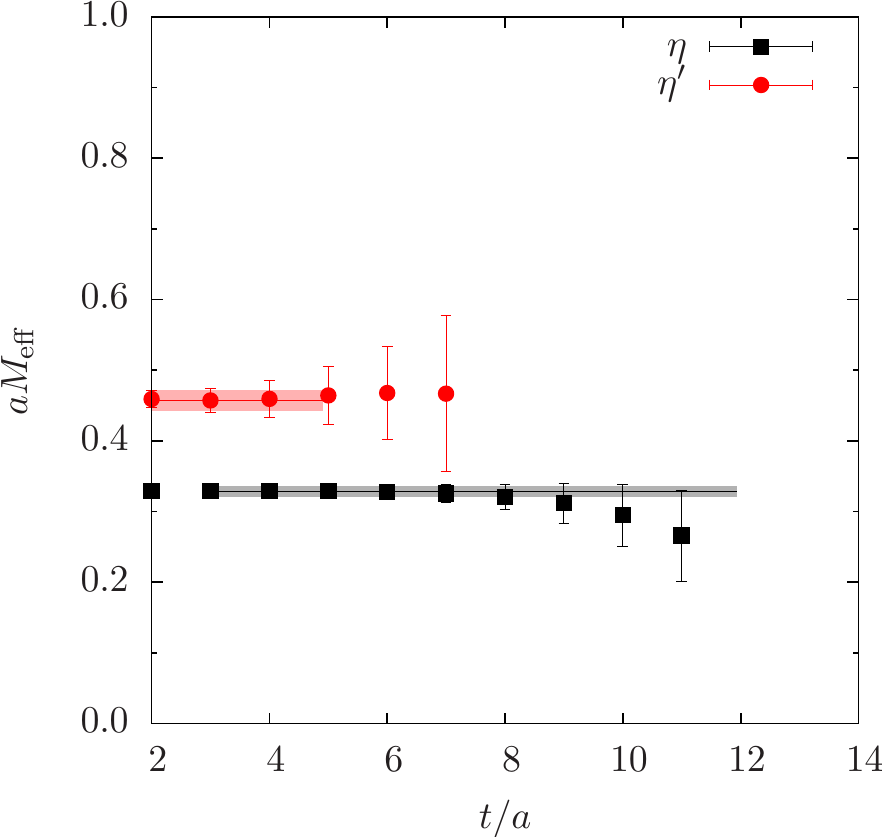}}\quad
  \subfigure[]{\includegraphics[width=.48\linewidth]{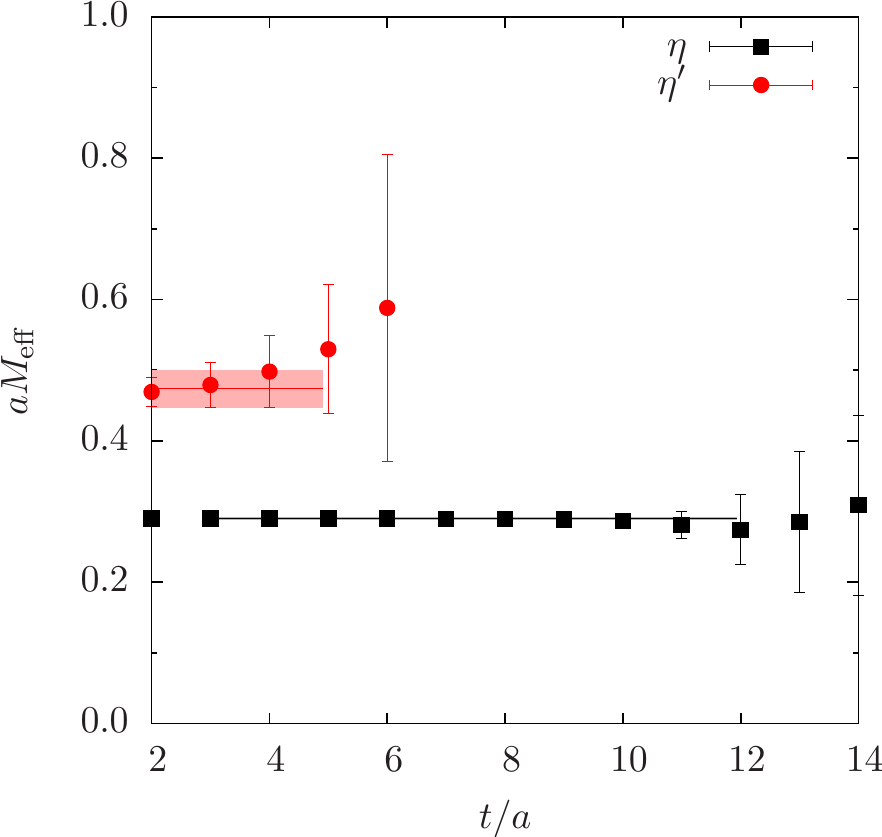}}
  \caption{ We show the effective masses $aM_\eta^\mathrm{OS}$ and
    $M_{\eta'}^\mathrm{OS}$ from a $3\times 3$ correlation matrix
    after subtraction of excited states as described in the text for
    ensemble $A60.24$. Panel (a) is for kaon matching and (b) for
    $\eta_s$  matching.}
  \label{fig:a60}
\end{figure}

In order to compare the mixed case with the unitary case we match the
two actions as detailed in the previous sections using either the kaon
or the $\eta_s$ mass. Next we compute OS meson masses at these
matching points. As an example we show in figure~\ref{fig:a60} the
effective masses for the principal correlators $\lambda^{(n)}(t,t_0)$ of
$\eta$ and $\eta'$ as a function of $t/a$ after removal of excited states
from the connected contributions.  For $\eta$ and
$\eta'$ a plateau in the effective masses is 
visible from early $t/a$ on. The corresponding result of an exponential
fit is indicated by the horizontal lines. The fit range corresponds to
the extension of the lines in $t/a$. All OS meson masses are compiled
in the appendix in tables~\ref{tab:results_M_eta} and
\ref{tab:results_M_etap}. 

The $\eta_c$ decouples from $\eta$ and $\eta'$ and the corresponding
signal is lost in noise very early in $t/a$. Hence, we will not
discuss it further here and due to the decoupling we will also not
discuss the charm quark mass dependence of operators in the following. 

It turns out that the choice of the matching variable makes a
significant difference for the extracted value of $M_\eta$.
Moreover, we always find $aM_\eta^{\eta_s} < aM_\eta^{K}$.
On the other hand, the value of $M_{\eta'}$ is unaffected within
statistical errors. We find this consistently for all the ensembles
investigated; cf. tables~\ref{tab:results_M_eta} and~\ref{tab:results_M_etap}.

In addition one observes $\phi^K<\phi^{\eta_s}$ by approximately
$15^\circ$; cf. table \ref{tab:results_phi}. This results from a
change in the overlap of mass and flavour eigenstates, leading to an
increased light quark contribution to the $\eta$ for kaon
matching. Consequently, the light quark contribution to the $\eta'$ is
reduced, while the respective strange quark contributions behave in
the opposite way. Since most of the noise is introduced by the light
quark disconnected diagrams in our calculation, this explains why
$M_\eta^{\eta_s}$ in general exhibits a smaller statistical error than
$M_\eta^K$, whereas the error for $M_{\eta'}^{\eta_s}$ is larger than
the one for $M_{\eta'}^K$. 

\begin{figure}[t]
 \centering
 \subfigure[]{\includegraphics[width=.48\linewidth]{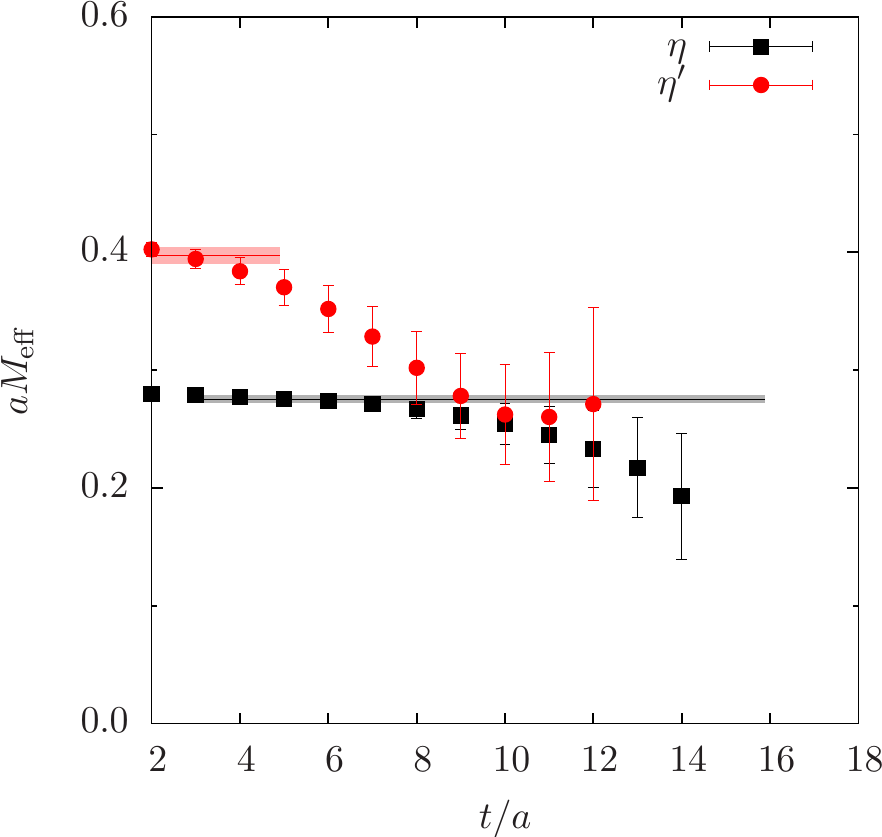}}\quad
 \subfigure[]{\includegraphics[width=.48\linewidth]{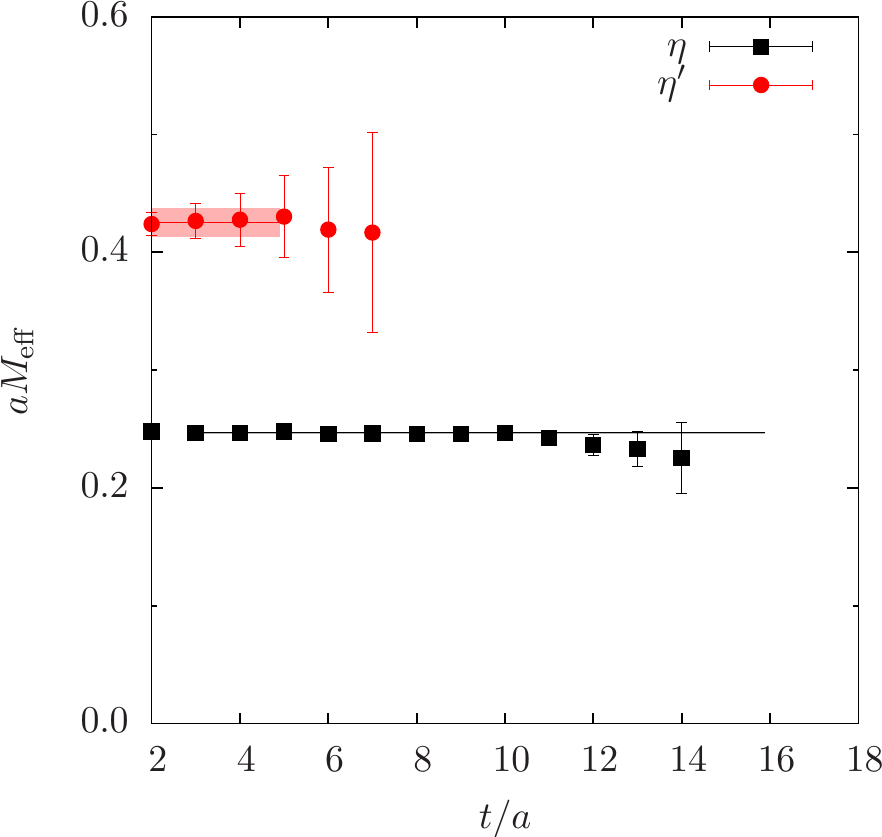}}
 \caption{We show the effective masses $aM_\eta$ and $M_{\eta'}$ of
   the two lowest-lying states on the $B55.32$ ensemble for (a) kaon
   matching and (b) the unitary case using subtraction of excited
   states as described in the text.} 
 \label{fig:b55}
\end{figure}

However, there is a tendency that kaon matching leads to worse
plateaus than $\eta_s$ matching. A particularly extreme case of this
behaviour is shown in figure~\ref{fig:b55} for the $B55.32$
ensemble. In the left panel the effective masses for the two lowest
lying states from kaon matching are plotted. Clearly there is no
plateau visible for the first excited state. For comparison and to
guide the eye we show the situation in the unitary setup in the right
panel, calculated on the same set of configurations. This is the only
ensemble for which we cannot identify a plateau for the $\eta'$
safely. Therefore, we will not quote a value for the $\eta'$ mass for
$B55.32$ and kaon matching.

Although the observed behaviour on $B55.32$ can still be interpreted as
a statistical fluctuation, it might -- in principle -- also be caused
by unitarity violation. However, it is neither possible to verify nor
exclude the latter from our present data. While earlier studies
\cite{Bowler:2004hs,Prelovsek:2004jp} observed a sign flip in the
scalar correlator signalling unitarity violation at least for a certain
regime of valence quark masses, a similar argument cannot easily be extended
to our case. The reason is that only the strange quark is treated in a
mixed action approach while the light quarks are unitary. When
looking at the scalar correlator made from strange quarks only, we do
not observe it to become negative on any of our ensembles.

Another observation regarding the two matching methods concerns the
behaviour of the ground states in the correlation functions used to
build the full correlation function matrix. One expects all
correlators with the same quantum numbers to asymptotically approach
the same ground state mass. We observe this for the unitary case,
where the $\eta$ mass can be extracted from all correlators in the
matrix $\mathcal{C}$ (diagonal and off-diagonal) at large Euclidean
times. In the OS case we observe a similar behaviour for $\eta_s$
matching, but for kaon matching e.g. the strange-strange correlator
alone does often not reproduce the $\eta$ mass from the light-light 
correlator. This might signal unitarity violations for the kaon
matching procedure on the one hand, and can explain the worse plateaus
for this particular choice of the matching observable on the other
hand. 

Finally, for $M_\eta^{\eta_s}$ we observe the error to be reduced
approximately by a factor of two with respect to the unitary result
for all ensembles (cf. table \ref{tab:results_M_eta}). We attribute
this to the fact that we can use the variance reduction method
discussed in section~\ref{sec:variance} for the OS strange quark,
which is not possible for the unitary strange quark. However, the
errors of $M_{\eta'}^{\eta_s}$ and $\phi^{\eta_s}$ do not show such an
error reduction (cf. table \ref{tab:results_M_etap} and
\ref{tab:results_phi}), presumably because the strange quark
contributes little to these observables. 

\subsection{Light Quark Mass Dependence}

The first goal of this paper is to compare unitary to mixed action
approaches and study the continuum limit of the corresponding
differences. For this purpose we will study differences of quantities
of the form $\Delta O = O^\mathrm{OS} - O^\mathrm{unitary}$. Our
ensembles at different values of the lattice spacing are not at
exactly identical light and strange sea quark masses. Therefore, we have
to understand whether we can nevertheless study the continuum limit.

Theoretically, the answer to this question is yes: both in the unitary
and in the mixed action approach we may write
\[
O^\mathrm{lat}\ = \ O^\mathrm{cont} + \mathcal{O}(a^2\Lambda_\mathrm{QCD}^2)
\]
and henceforth the aforementioned difference $\Delta O$ is
$\mathcal{O}(a^2\Lambda_\mathrm{QCD}^2)$. A quark mass dependence 
is expected to be negligible because for the quark mass difference 
$\delta\mu \ll\Lambda_\mathrm{QCD}$ holds. Since $\Delta O$ is always
computed on identical gauge configurations there is no physical quark
mass dependence that needs to be taken into account, because it
cancels in the difference.

\begin{figure}[t]
  \centering
  \subfigure[]{\includegraphics[width=0.45\linewidth]{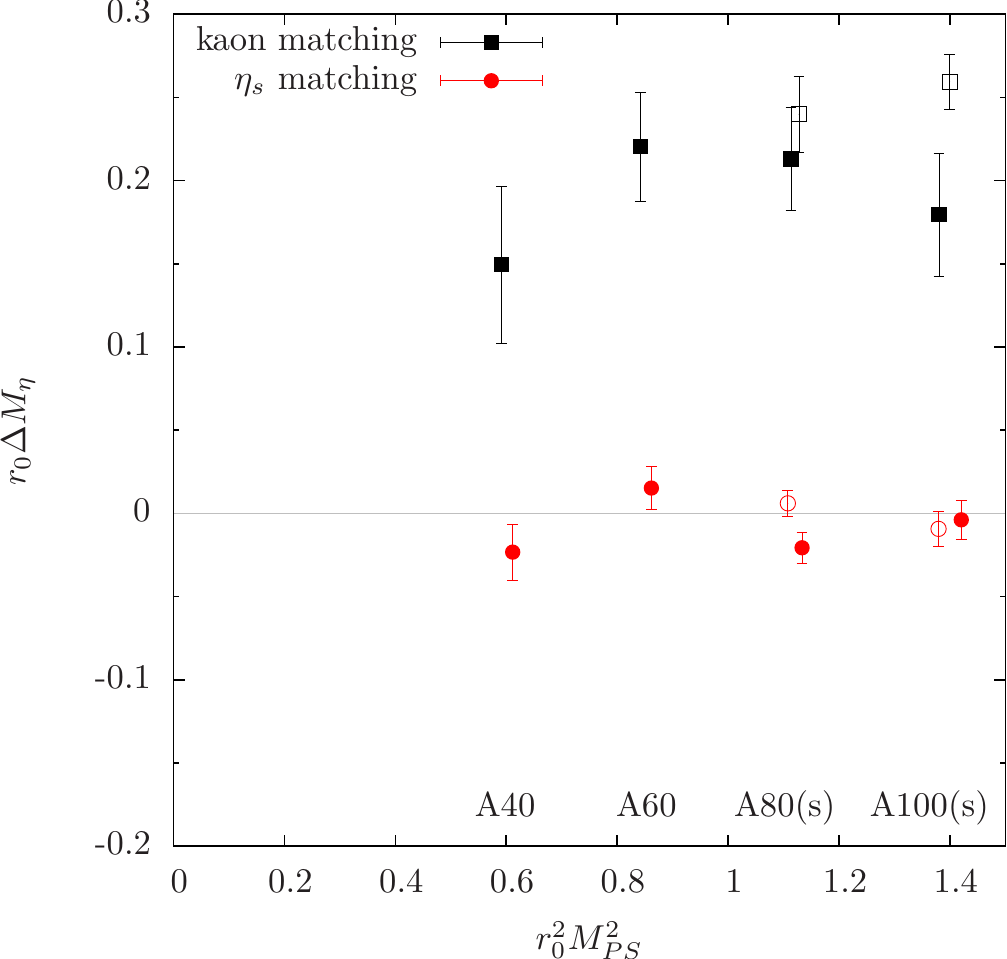}}\quad
  \subfigure[]{\includegraphics[width=0.45\linewidth]{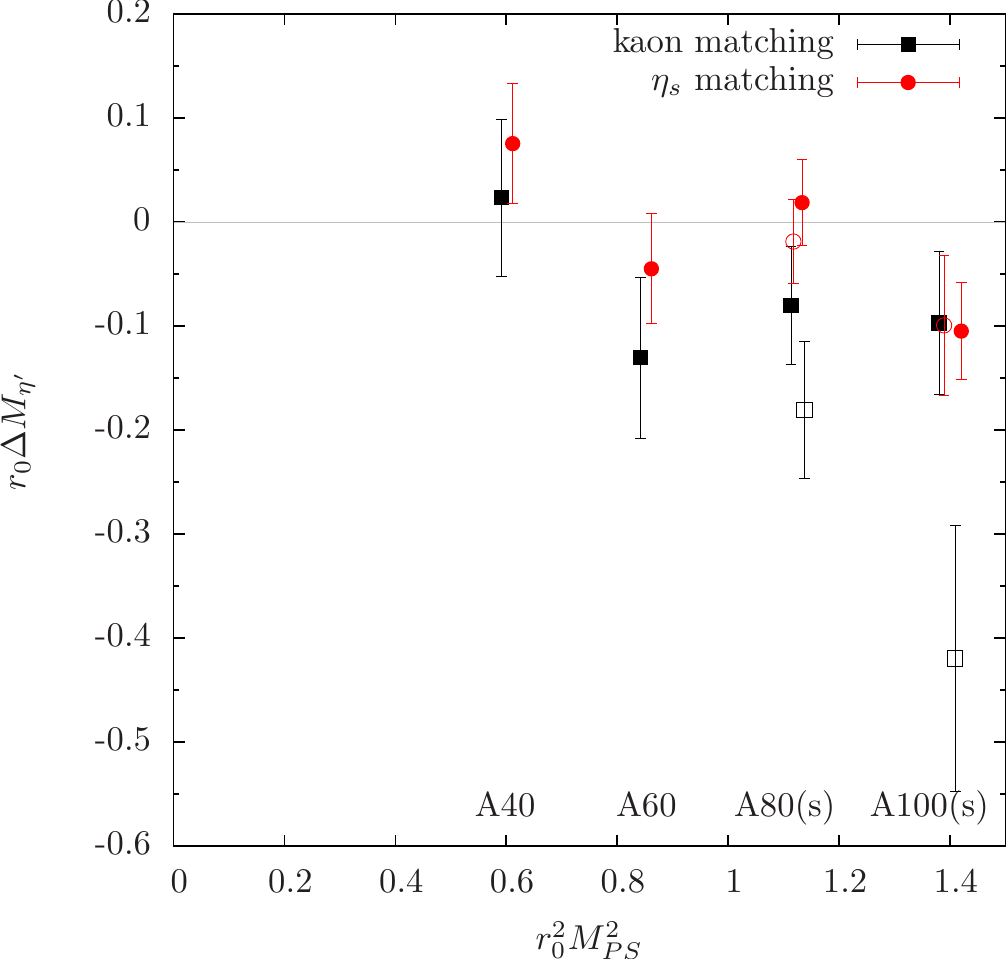}}
  \caption{(a) $r_0\Delta M_\eta$ as a function of $(r_0\mps)^2$ for
    the four A ensembles A40, A60, A80(s) and A100(s). Open symbols
    correspond to the s-ensembles. (b) like (a), but for
    $r_0\Delta M_{\eta'}$. The kaon matching and s-ensemble data have
    been displaced horizontally for better legibility.}
  \label{fig:mPsDiffDep}
\end{figure}

Despite this theoretical argument, let us also investigate this point
numerically. We first study the light quark mass dependence of the
difference between unitary and OS values of $M_\eta$ and
$M_{\eta'}$. For this purpose we focus on the A-ensembles A40, A60,
A80 and A100, where we have different light quark mass values
available. We denote
\[
\Delta M_X = M_X^\mathrm{OS} - M_X^\mathrm{unitary}\,,\quad X=\eta,\eta'
\]
the difference between unitary and OS meson masses. Analogously we
define the angle difference $\Delta\phi$. $r_0 \Delta M_\eta$ is shown
for the A-ensembles in the left panel of 
figure~\ref{fig:mPsDiffDep} as a function of $(r_0\mps)^2$. The
chirally extrapolated values of $r_0/a$ used in this study have been
determined in \cite{Carrasco:2014cwa} and are listed in
table~\ref{tab:r0}. 
Filled circles represent the $\eta_s$ matching results, filled boxes the
corresponding kaon matching results. The differences are computed
using exactly the same configurations leading to reduced statistical
errors due to the correlation between unitary and OS data.

\begin{table}[t!]
 \centering
 \begin{tabular*}{.6\textwidth}{@{\extracolsep{\fill}}lccc}
  \hline\hline
   $\beta$      & $1.90$ & $1.95$ & $2.10$ \\
   $r_0/a$ & $5.31(8)$ & $5.77(6)$ & $7.60(8)$ \\
  \hline\hline 
 \end{tabular*}
 \caption{The chirally extrapolated values for $r_0/a$ at each value
   of $\beta$ corresponding to the three different lattice spacings
   \cite{Carrasco:2014cwa}.}
 \label{tab:r0}
\end{table}

For $\eta_s$ matching $r_0 \Delta M_\eta$ is for all four investigated
ensembles compatible with zero within one sigma, while for kaon
matching the difference is always positive and not compatible with
zero. For both matching procedures, but in particular for $\eta_s$
matching, the dependence on $(r_0\mps)^2$ is not significant within
our statistical uncertainties.

In the right panel of figure~\ref{fig:mPsDiffDep} we show $r_0 \Delta
M_{\eta'}$ as a function of $(r_0\mps)^2$. Despite the larger
uncertainties, the differences are compatible with zero for all
ensembles and both matching procedures. There is a slight trend for
differences with larger modulus for kaon matching. Like for the
$\eta$ the light quark mass dependence is not significant. 

The angle difference $\Delta\phi$ shows a very similar behaviour to
$\Delta M_\eta$, see tables~\ref{tab:resultsDiffK} and
\ref{tab:resultsDiffetas}. For $\eta_s$ matching the difference is
compatible with zero, while for kaon matching a value of about $-15^\circ$ is
observed. Also here the light quark mass dependence is not
significant. Besides, we find that the difference between $\phi_\ell$ and
$\phi_s$ is compatible with zero for both matching methods and compatible to the
one found in the unitary setup~\cite{Michael:2013gka}, again
confirming the smallness of OZI suppressed corrections. 

In order to check the strange quark mass dependence of the differences
$\Delta M$ and $\Delta\phi$ we make use of the A80, A80s and A100,
A100s ensembles. The s-ensembles differ from their non-s counterparts
only by a different bare strange quark mass value. The corresponding values
for the differences defined before are also displayed in
figure~\ref{fig:mPsDiffDep} with open symbols. For $\eta_s$ matching
the differences show no dependence on the strange quark mass, whereas
this cannot be concluded completely for kaon matching. In particular,
we see for A100s deviations for kaon matching, but statistical errors
can still account for the deviation. 

\subsection{Continuum Limit}

Next, we study the dependence on the lattice spacing. For this purpose
we use the ensembles $A60.24$, $B55.32$ and $D45.32sc$, which have approximately the
same physical value of the pion mass, i.e. $r_0 M_\mathrm{PS}^{A60.24}
= 0.917(14)_\mathrm{stat}$, $r_0 M_\mathrm{PS}^{B55.32} =
0.888(09)_\mathrm{stat}$ and $r_0 M_\mathrm{PS}^{D45.32sc} =
0.911(11)_\mathrm{stat}$,  
where we included the statistical error from the respective, chirally extrapolated values of $r_0/a$. As discussed in the previous
section, we do not expect the residual differences in the light and
strange quark masses to have any effect on this study.

\begin{figure}[t]
  \centering
  \subfigure[]{\includegraphics[width=0.48\textwidth]{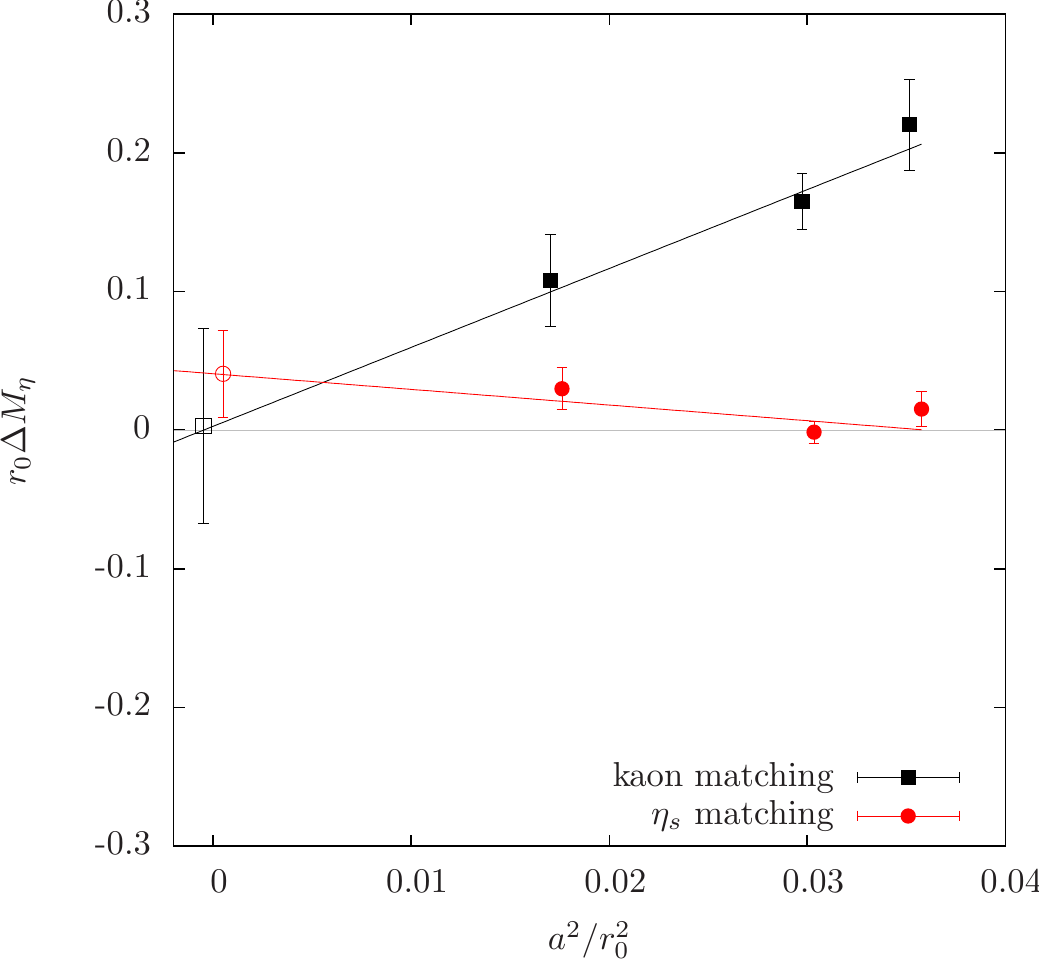}}
  \subfigure[]{\includegraphics[width=0.48\textwidth]{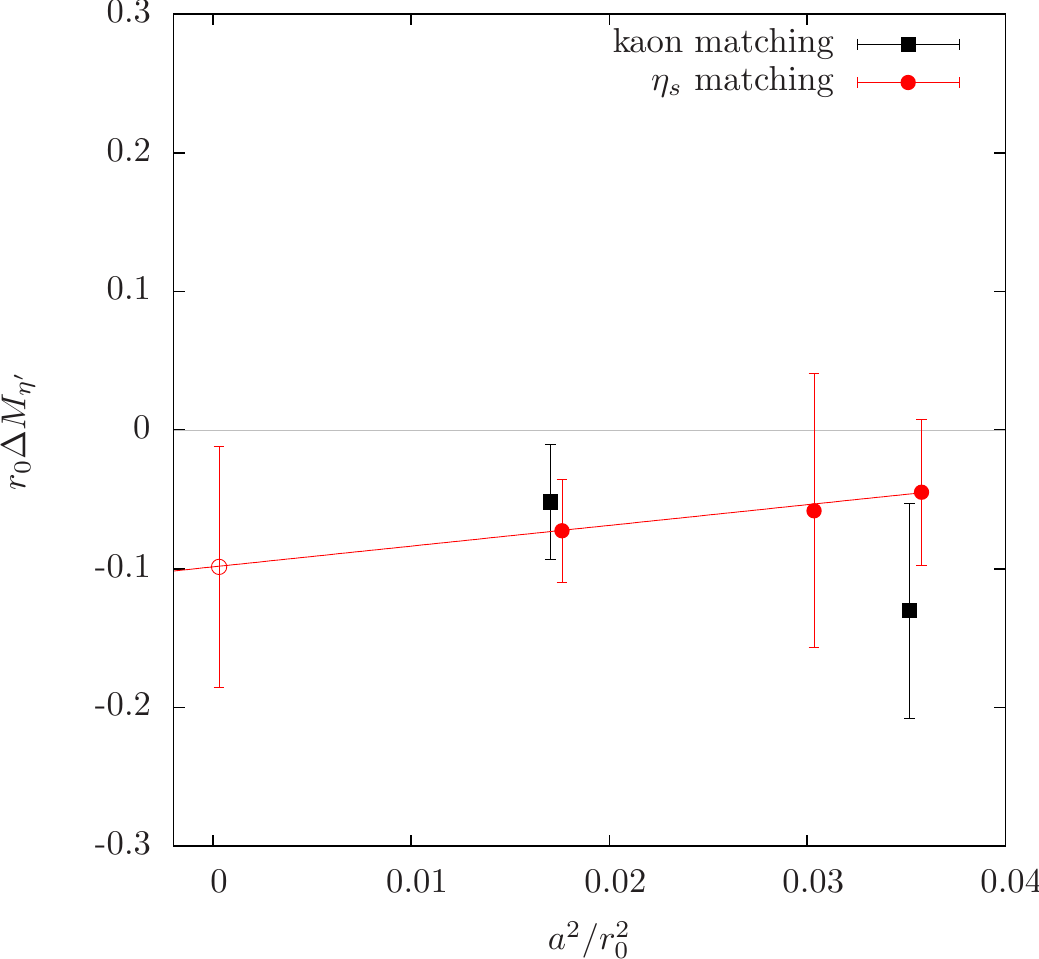}}
  \caption{Continuum extrapolation of (a) $r_0\Delta M_{\eta}$ and (b)
    $r_0\Delta M_{\eta'}$ as a function of $(r_0/a)^2$. We show the
    continuum extrapolated results from a linear extrapolation of the
    three ensembles ($D45.32sc$, $B55.32$, $A60.24$) in $(a/r_0)^2$ as open
    symbols. The 
    continuum results are displaced horizontally for legibility.}
  \label{fig:aDepM}
\end{figure}

The difference between OS and unitary results $\Delta M$ for both
matching procedures is shown for $\eta$ and $\eta'$ in
figure~\ref{fig:aDepM} as a function of $(a/r_0)^2$ in the left and
right panel, respectively. The lines represent linear fits in
$(a/r_0)^2$ to our data, and the corresponding continuum extrapolated
values are shown with open symbols. 

$r_0\Delta M_\eta$ is shown in the left panel of
figure~\ref{fig:aDepM}. For both
matching procedures we observe a linear dependence in $(a/r_0)^2$. A
corresponding continuum extrapolation in $(a/r_0)^2$ leads to the
expected vanishing of this difference at $a=0$ within errors.
Kaon matching clearly exhibits larger $a^2$ artifacts, while $\eta_s$ matching gives
$r_0\Delta M_\eta$ compatible with zero for each value of the lattice
spacing separately. 

In the right panel of figure~\ref{fig:aDepM} we
show $\Delta M_{\eta'}$, again for both matching procedures. We remark
that for kaon matching it is not possible to perform a fit from our
present data, because of the missing mass value on the $B55.32$
ensemble which is due to a bad plateau, as discussed above. 
In this case, statistical errors are significantly larger. However, within
their larger errors the difference for the two matching procedures
seems compatible and the difference vanishes in the continuum limit for
$\eta_s$ matching, as indicated by the fitted line in the
plot. In contrast to the $\eta$ mass, 
it cannot be concluded that lattice artifacts for kaon matching
are significantly larger than for $\eta_s$ matching.

\begin{figure}[t]
  \centering
  \subfigure[]{\includegraphics[width=0.47\linewidth]{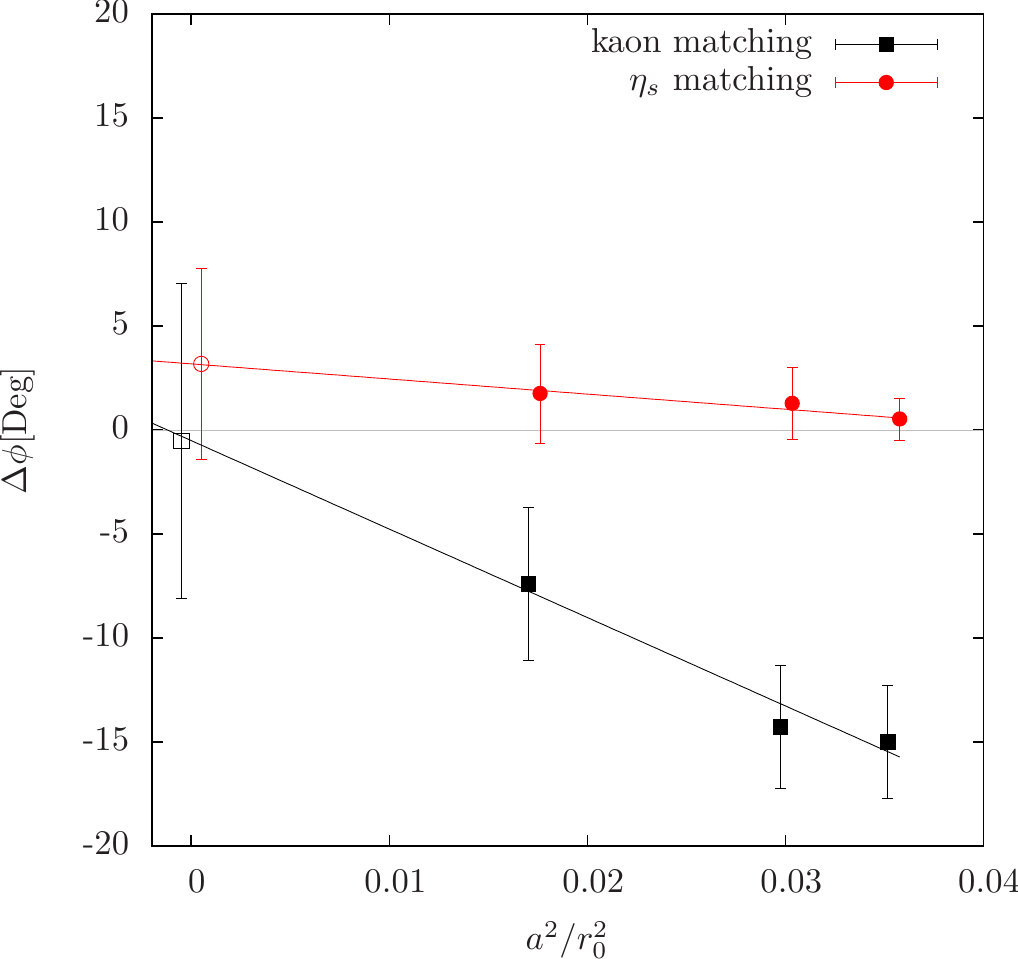}}
  \subfigure[]{\includegraphics[width=0.49\linewidth]{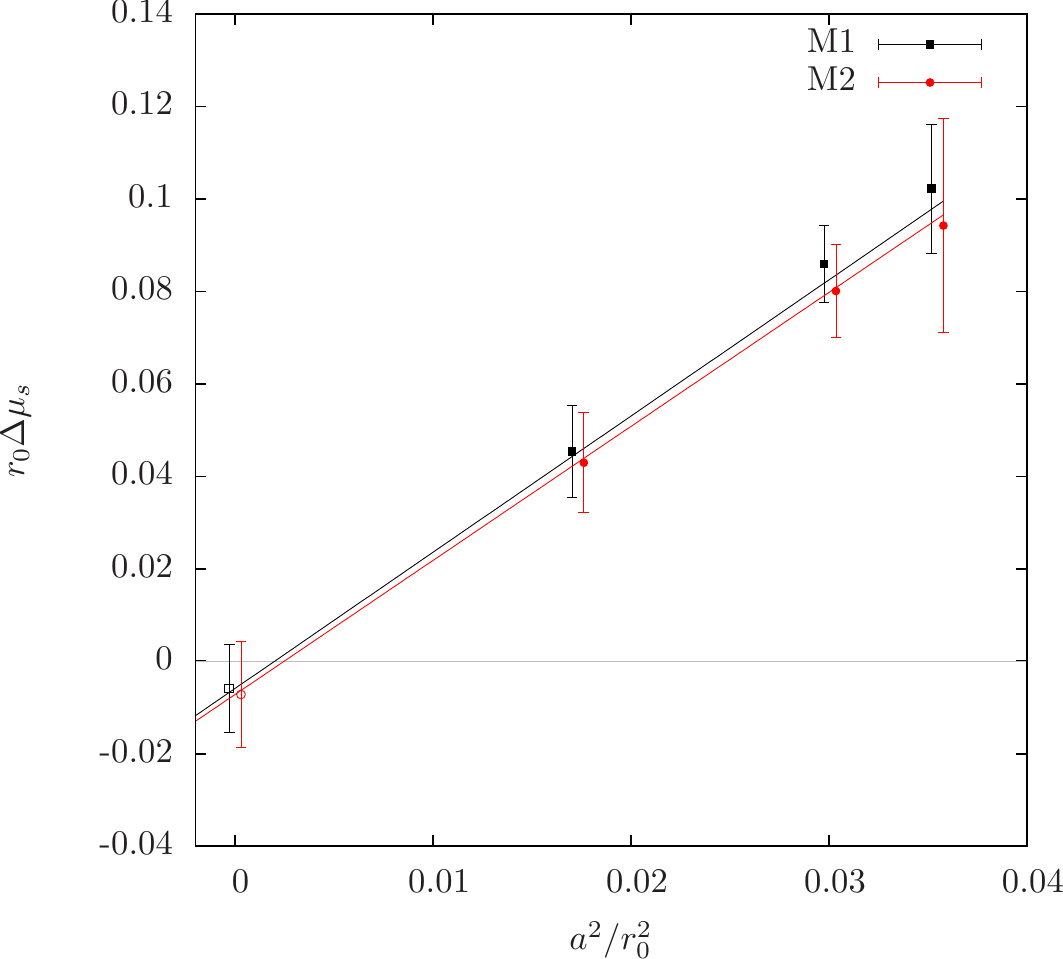}}
  \caption{(a) like figure~\ref{fig:aDepM}, but for the mixing
    angle  difference $\Delta\phi$. In (b) we show $r_0\Delta\mu_s$
    at $2\ \mathrm{GeV}$ in the $\overline{\mathrm{MS}}$ scheme as a function
    of $(a/r_0)^2$ for the two methods M1 and M2 to estimate $Z_P$
    presented in Ref.~\cite{Carrasco:2014cwa}.}
  \label{fig:aDepAngle}
\end{figure}

In the left panel of figure~\ref{fig:aDepAngle} $\Delta\phi$ is shown
as a function of $(a/r_0)^2$, again for the ensembles $A60.24$, $B55.32$ and
$D45.32sc$. Like for $\Delta M_\eta$ we observe also for $\Delta\phi$ larger
differences for kaon matching compared to $\eta_s$ matching. For $\eta_s$ matching the
difference is in fact compatible with zero for all three ensembles
separately. For both matching procedures the continuum extrapolated
values are compatible with zero.

Finally, we show in the right panel of figure~\ref{fig:aDepAngle} the
quark mass difference 
\begin{equation}
  \label{eq:dmu}
  \Delta\mu_s = \frac{1}{Z_P}(\mu_s^{K} - \mu_s^{\eta_s})
\end{equation}
as a function of $(a/r_0)^2$ for the three ensembles $A60.24$, $B55.32$ and
$D45.32sc$ at $2\ \mathrm{GeV}$ in the $\overline{\mathrm{MS}}$ scheme. The
renormalisation constant $Z_P$ has been taken from
Ref.~\cite{Carrasco:2014cwa}. The two colours correspond to the
methods M1 and M2 for estimating $Z_P$. We refer to
Ref.~\cite{Carrasco:2014cwa} for the details. 
In the continuum limit it is expected
that the two matching conditions agree and the difference should
vanish like $a^2$. This is what is confirmed by
figure~\ref{fig:aDepAngle} (b). Also, the two methods M1 and M2 give
compatible results in the continuum limit, as expected.

\subsection{Dependence on Sea and Valence Strange Quark Mass}

\begin{table}[t!]
  \centering
  \begin{tabular*}{.6\textwidth}{@{\extracolsep{\fill}}l r}
    \hline\hline
    ensemble   & $D_\eta^\mathrm{val}$\\
    \hline\hline
    $A40.24$   & $0.76(22)$ \\
    $A60.24$   & $0.97(17)$ \\
    $A80.24$   & $1.17(17)$ \\
    $A100.24$  & $0.92(18)$ \\
    \hline
    $A80.24s$  & $1.07(11)$\\
    $A100.24s$ & $1.32(08)$ \\
    \hline
    $B55.32$   & $0.93(09)$ \\
    $D45.32sc$   & $0.86(31)$ \\
    \hline\hline
  \end{tabular*}
  \caption{The valence derivative $D_\eta^\mathrm{val}$ obtained in
    the OS case by using the mass values from the kaon and $\eta_s$
    matching points.}
  \label{tab:resultsDEta}
\end{table}

\begin{table}[t!]
  \centering
  \begin{tabular*}{.6\textwidth}{@{\extracolsep{\fill}}l rr}
    \hline\hline
    ensemble   & $D_\eta$ & $D_\eta^\mathrm{OS}$ \\
    \hline\hline
    A80/A80s   & $1.54(13)$ & $1.37(07)$ \\
    A100/A100s & $1.34(15)$ & $1.67(11)$ \\
    \hline\hline
  \end{tabular*}
  \caption{We list the values for $D_\eta$ evaluated both for the
    unitary and the OS case using the two A80 and the two A100
    ensembles. This derivative includes, in contrast to
    $D_\eta^\mathrm{val}$, the valence and sea strange quark mass
    dependence. The values for $D_\eta^\mathrm{OS}$ are for $\eta_s$
    matching}
  \label{tab:resultsDfull}
\end{table}

\begin{figure}[t!]
  \centering
  \includegraphics[width=.7\linewidth]{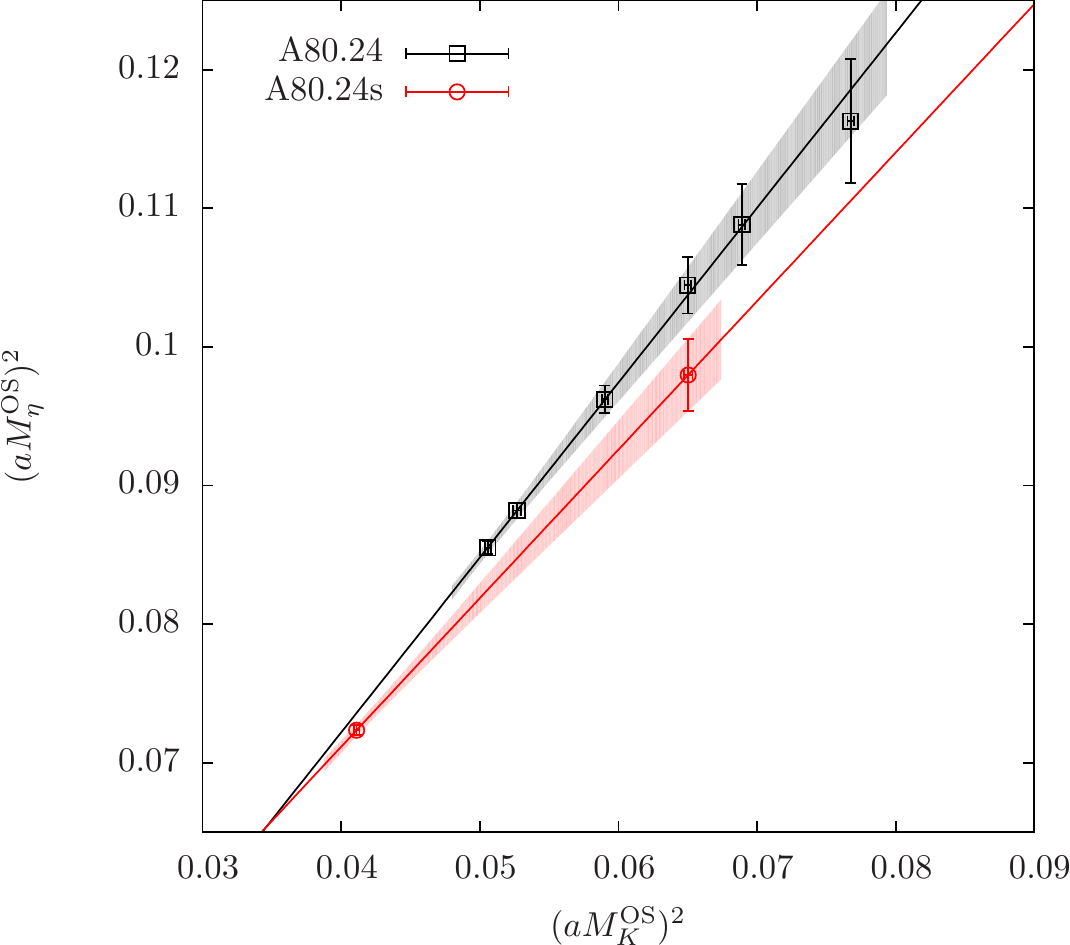}
  \caption{$(M_{\eta}^\mathrm{OS})^2$ as a function of
    $(M_{K}^\mathrm{OS})^2$ for ensembles  A80.24 and A80.24s.}
  \label{fig:Metasq}
\end{figure}

Next we study the dependence of the $\eta$ (and in principle also the
$\eta'$) meson mass on the valence and sea quark mass values. As said
in the introduction, the dependence on the valence and sea quark
masses must be identical (at least within errors) to legitimate
re-tuning in the valence quark masses only against sea strange quark mass mismatches. For this purpose
we first define the dimensionless quantity
\begin{equation}
  \label{eq:DetaVal}
  D_\eta^\mathrm{val}\ = \ \left. \frac{d(M_\eta^\mathrm{OS})^2}{d(M_{K}^\mathrm{OS})^2} \right|_\mathrm{fixed\ sea\ ensemble}\,,
\end{equation}
which can be computed using the two matching points we have available
for each ensemble. For estimating $D_\eta^\mathrm{val}$ from two
$\mu_s$ values at each ensemble, we have to assume that
$(M_\eta^\mathrm{OS})^2$ depends linearly on $(M_{K}^\mathrm{OS})^2$ to a good
approximation. That this is the case can be seen in
figure~\ref{fig:Metasq}, where we show $(M_\eta^\mathrm{OS})^2$ as a
function of ($M_K^\mathrm{OS})^2$ for the two ensembles A80.24 and
A80.24s. The lines represent linear fits to our data, which describe
the data well within errors. 

We expect $D_\eta^\mathrm{val}$ to be mostly sensitive to
the valence strange quark mass if computed for several valence
$\mu_s$-values on the same ensemble. The results for
$D_\eta^\mathrm{val}$ are compiled in 
table~\ref{tab:resultsDEta} for all ensembles and $\eta_s$
matching. They appear -- independently of the lattice spacing, light
and strange quark mass values -- to be all compatible with $1$. Taking
the weighted average we obtain $D_\eta^\mathrm{val}=1.09(5)$, where
the error is purely statistical. Concerning possible systematics we
stress that there is no trend visible from the data, e.g. regarding a
quark mass or lattice spacing dependence. 

Including the sea strange quark mass dependence, the corresponding
derivative is given by
\[
D_\eta\ = \ \frac{d(M_\eta)^2}{d(M_{K})^2}\,.
\]
For the unitary case we find $D_\eta=1.45(10)$ and in the OS case
$D_\eta^\mathrm{OS}=1.46(6)$, using the A100.24, A100.24s and A80.24, 
A80.24s ensembles and $\eta_s$ matching, see
table~\ref{tab:resultsDfull}. While $D_\eta$ and $D_\eta^\mathrm{OS}$
are compatible within errors, they differ significantly from
$D_\eta^\mathrm{val}$. For kaon matching the relative statistical errors on $D_\eta^\mathrm{OS}$ 
turn out to be at least a factor five larger than for $\eta_s$ matching. 
Therefore, a meaningful statement regarding the compatibility of $D_\eta^\mathrm{OS}$ 
with $D_\eta$ and $D_\eta^\mathrm{val}$ for kaon matching is not 
possible from our current data.

We take this as an indication that $M_\eta$ is indeed a quantity with
a significant sea strange quark mass dependence. Therefore, correcting
for mismatches of the sea strange quark mass value in the valence
sector only is not enough for $M_\eta$. 

In principle, the difference between $D_\eta$ and
$D_\eta^\mathrm{val}$ that we found for a single lattice spacing could
also be a 
lattice artifact. We do not think this is the case for two reasons:
first, in Ref.~\cite{Ottnad:2012fv} the value of $D_\eta$ was used to
correct a mismatch in the strange quark mass tuning for all three
lattice spacings available. And we did not observe large cut-off
effects introduced by this procedure. Second, also
$D_\eta^\mathrm{OS}$ is merely independent of the lattice spacing. 
In fact, as the $D_X$ are computed from differences and the leading
lattice artifacts are independent of the quark mass, it is expected
that these quantities are not plagued by large cut-off effects.

Unfortunately, the statistical uncertainty on the $\eta'$ meson masses
is too large to allow for a meaningful investigation of
$D_{\eta'}$. Within errors this quantity is always zero, irrespective
of whether the valence or the full strange quark mass dependence is
considered. Moreover, $M_{\eta'}$ has a larger light than strange
quark contribution. It would, therefore, be interesting to perform the
same study for the valence light quark mass instead of the valence
strange. 

Finally, we remark that it is in principle possible to calculate the
difference between $D_\eta$ and $D_\eta^\mathrm{val}$ from chiral
perturbation
theory~\cite{Bernard:1992mk,Sharpe:2000bc,Bijnens:2006jv}. At leading
order the corresponding prediction is $4/3$ for both derivatives,
implying that the difference is an NLO effect.

\section{Summary and Discussion}

In this paper we have studied $\eta$ and $\eta'$ mesons in a mixed
action approach and in comparison to the unitary results. The mixed action
was so-called Osterwalder-Seiler fermions on a twisted mass sea with
$N_f=2+1+1$ dynamical quark flavours. 

We have found that indeed the difference between mixed and unitary
results vanishes as the continuum limit is approached. The rate is as
expected of $\mathcal{O}(a^2)$~\cite{Frezzotti:2004wz} for all
quantities and matching procedures investigated in this paper. 

For the $\eta$ mass we find a significant dependence of
the size of the cutoff effects on the matching procedure. Lattice
artifacts in the difference to the unitary result are compatible with
zero when the two actions are matched using the $\eta_s$ meson, while
they are of normal size when the kaon is used as a matching
variable. The same is true for the mixing angle and the strange quark
mass. 

For $M_{\eta'}$ we do not observe a strong dependence on the matching
procedure. This can have two reasons: first the error of $M_{\eta'}$ is
large making precise statements difficult. Second, the $\eta'$
receives a strong contribution from sea quarks, because it is mainly
the singlet state. The sea quark contributions are unaffected by
different choices of the valence strange quark mass. Hence, this
finding might reflect the physical properties of the $\eta'$ meson. 

This shows that the mixed action approach can also be applied in
practice for flavour singlet quantities and, more generally, for
observables involving fermionic disconnected diagrams. In case of
$\eta_s$ matching we find even reduced statistical errors for the
$\eta$ mass which might turn out to be an important advantage of the
mixed approach. Thus, we will use the mixed action to investigate more
complicated problems like $\eta\to\gamma\gamma$ form factors or $K\pi$
scattering for $I=1/2$ in the future.

Another important result of this paper is that for the $\eta$ meson it is 
not sufficient to re-tune the valence quark masses to correct for small
mismatches in the simulation runs. Our data shows that the valence
strange quark mass dependence of the $\eta$ differs significantly from
the dependence on the sea plus valence strange quark mass. And this
difference is not vanishing as the continuum limit is approached. For
the $\eta'$ we cannot make such a statement due to too large statistical
errors, but we expect a similar result once $M_{\eta'}$ can be
determined with higher accuracy.

The latter finding seems to contradict previous studies where the
valence quark masses have been re-tuned to their physical
values in the same mixed action approach for instance to determine
non-singlet pseudo-scalar decay constants or quark masses (see for
instance
Refs.~\cite{Carrasco:2014cwa,Carrasco:2014poa,Blossier:2009bx}).
However, these investigations were concerned with observables 
for which the quark mass dependence is expected to be mainly governed
by the valence quarks. For the $\eta$ and $\eta'$ mesons studied here
this is not the case as OZI violating contributions are anomalously
large. However, with high enough accuracy the effect seen here should
also show up in other physical quantities, but on the current level of
precision it is likely to be a negligible systematic uncertainty.

\subsection*{Acknowledgements} 

We thank the
members of ETMC for the most enjoyable collaboration. The computer
time for this project was made available to us by the John von
Neumann Institute for Computing (NIC) on the JUDGE, Jugene and Juqueen
systems in J{\"u}lich. In
particular we thank U.-G.~Mei{\ss}ner for granting us 
access on JUDGE. We thank C.~Michael and S.~Simula for useful comments.
We thank R.~Frezzotti for useful discussions and very helpful comments
on the draft of this paper.
This project was funded by the DFG as a project in
the SFB/TR 16 and in part by the Sino-German CRC110. Two of the
authors (K.O. and C.U.) were supported by the Bonn-Cologne Graduate
School (BCGS) of Physics and Astronomy. The open source software
packages tmLQCD~\cite{Jansen:2009xp}, Lemon~\cite{Deuzeman:2011wz} and
R~\cite{R:2005} have been used.

\bibliographystyle{elsarticle-num}
\bibliography{bibliography}

\newpage
\begin{appendix}
  \section{Data Tables}
\label{sec:app}

In this appendix we have compiled all data in tables for convenience. 

\begin{table}[h!]
 \centering
 \begin{tabular*}{1.\textwidth}{@{\extracolsep{\fill}}lcccc}
  \hline\hline
  ensemble & $aM_\eta$ & $aM_\eta^{\eta_s}$ & $aM_\eta^{K}$ \\
  \hline\hline
  $A40.24$   & $0.2837(47)$ & $0.2793(35)$ & $0.312 (12)$ \\
  $A60.24$   & $0.2870(28)$ & $0.2899(16)$ & $0.3285(74)$ \\
  $A80.24$   & $0.3009(20)$ & $0.2970(09)$ & $0.3410(67)$ \\
  $A100.24$  & $0.3074(23)$ & $0.3067(13)$ & $0.3412(75)$ \\
  \hline
  $A80.24s$  & $0.2678(13)$ & $0.2690(07)$ & $0.3130(42)$ \\
  $A100.24s$ & $0.2759(17)$ & $0.2741(10)$ & $0.3247(30)$ \\
  \hline
  $B55.32$   & $0.2467(12)$ & $0.2465(07)$ & $0.2753(34)$ \\
  $D45.32sc$ & $0.1890(34)$ & $0.1929(39)$ & $0.2032(72)$ \\
  \hline\hline
 \end{tabular*}
 \caption{Results for $aM_\eta$ for the unitary and the mixed action approach. For the latter we show the values corresponding to $\eta_s$ and kaon matching.}
 \label{tab:results_M_eta}
\end{table}

\begin{table}[h!]
 \centering
 \begin{tabular*}{1.\textwidth}{@{\extracolsep{\fill}}lcccc}
  \hline\hline
  ensemble & $aM_{\eta'}$ & $aM_{\eta'}^{\eta_s}$ & $aM_{\eta'}^{K}$ \\
  \hline\hline
  $A40.24$   & $0.443(27)$ & $0.457(28)$ & $0.448(15)$ \\
  $A60.24$   & $0.482(27)$ & $0.474(27)$ & $0.458(15)$ \\
  $A80.24$   & $0.481(26)$ & $0.485(29)$ & $0.466(17)$ \\
  $A100.24$  & $0.461(24)$ & $0.441(21)$ & $0.442(13)$ \\
  \hline
  $A80.24s$  & $0.465(23)$ & $0.461(25)$ & $0.431(13)$ \\
  $A100.24s$ & $0.542(42)$ & $0.523(40)$ & $0.463(20)$ \\
  \hline
  $B55.32$   & $0.425(12)$ & $0.415(12)$ & bad plateau \\
  $D45.32sc$ & $0.278(12)$ & $0.269(12)$ & $0.271(09)$ \\
  \hline\hline
 \end{tabular*}
 \caption{Same as table~\ref{tab:results_M_eta}, but for $aM_{\eta'}$.}
 \label{tab:results_M_etap}
\end{table}

\begin{table}[h!]
 \centering
 \begin{tabular*}{1.\textwidth}{@{\extracolsep{\fill}}lcccc}
  \hline\hline
  ensemble & $\phi$ & $\phi^{\eta_s}$ & $\phi^{K}$ \\
  \hline\hline
  $A40.24$   & $47.1(2.0)$ & $47.3(2.2)$ & $30.6(4.1)$ \\
  $A60.24$   & $49.2(1.4)$ & $49.8(1.6)$ & $34.2(3.7)$ \\
  $A80.24$   & $49.9(1.1)$ & $51.1(1.1)$ & $38.0(3.5)$ \\
  $A100.24$  & $49.6(1.3)$ & $50.3(1.3)$ & $31.9(4.1)$ \\
  \hline
  $A80.24s$  & $51.3(0.8)$ & $52.6(0.5)$ & $36.2(3.2)$ \\
  $A100.24s$ & $53.8(0.7)$ & $55.1(0.6)$ & $44.2(2.7)$ \\
  \hline
  $B55.32$   & $48.2(0.8)$ & $49.5(1.3)$ & $34.0(2.9)$ \\
  $D45.32sc$ & $45.6(2.5)$ & $47.3(4.4)$ & $38.2(5.8)$ \\
  \hline\hline
 \end{tabular*}
 \caption{Same as table~\ref{tab:results_M_eta}, but for the mixing angle $\phi$.}
 \label{tab:results_phi}
\end{table}

\begin{table}[h!]
 \centering
 \begin{tabular*}{.8\textwidth}{@{\extracolsep{\fill}}lccc}
  \hline\hline
  ensemble &  $a\Delta M_\eta$ &  $a\Delta M_{\eta'}$ &  $\Delta \phi[Deg]$  \\ 
  \hline\hline
  $A40.24$        &$+0.0281(89)$  &$+0.004(14)$  & $-16.5(2.6)$ \\
  $A60.24$        &$+0.0415(61)$  &$-0.024(15)$  & $-15.0(2.7)$ \\
  $A80.24$        &$+0.0401(58)$  &$-0.015(11)$  & $-11.9(2.6)$ \\
  $A100.24$       &$+0.0338(69)$  &$-0.018(13)$  & $-17.7(3.3)$ \\
  \hline                                                             
  $A80.24s$       &$+0.0452(43)$  &$-0.034(12)$  & $-15.1(2.8)$ \\
  $A100.24s$      &$+0.0489(31)$  &$-0.079(24)$  & $-9.5(2.7) $ \\
  \hline                                                               
  $B55.32$        &$+0.0286(35)$  &$N/A$         & $-14.3(2.9)$ \\
  $D45.32sc$      &$+0.0142(44)$  &$-0.007(06)$  & $-7.4(3.7) $ \\
  \hline\hline
 \end{tabular*}
 \caption{Values for  $a\Delta M_\eta$,  $a\Delta
   M_{\eta'}$ and $\Delta \phi[Deg]$ as defined in the main text for
   kaon matching procedure.}
 \label{tab:resultsDiffK}
\end{table}

\begin{table}[h!]
 \centering
 \begin{tabular*}{.8\textwidth}{@{\extracolsep{\fill}}lccc}
  \hline\hline
  ensemble &  $a\Delta M_\eta$ &  $a\Delta M_{\eta'}$ &  $\Delta \phi[Deg]$  \\ 
  \hline\hline
  $A40.24$     &$-0.0044(32)$  &$+0.014(11)$  & $+0.2(1.2) $ \\
  $A60.24$     &$+0.0029(24)$  &$-0.008(10)$  & $+0.5(1.0)  $ \\
  $A80.24$     &$-0.0039(18)$  &$+0.004(08)$  & $-1.2(0.6) $ \\
  $A100.24$    &$-0.0007(22)$  &$-0.020(09)$  & $+0.7(1.0) $ \\
  \hline                                                             
  $A80.24s$    &$+0.0011(15)$  &$-0.004(08)$  & $+1.3(0.9) $ \\
  $A100.24s$   &$-0.0018(20)$  &$-0.019(13)$  & $+1.3(0.8)  $ \\
  \hline                                                               
  $B55.32$     &$-0.0002(14)$  &$-0.010(17)$  & $+1.3(1.5)  $ \\
  $D45.32sc$   &$+0.0039(20)$  &$-0.010(05)$  & $+1.8(2.4)  $ \\
  \hline\hline
 \end{tabular*}
 \caption{Same as table~\ref{tab:resultsDiffK}, but for $\eta_s$ matching procedure.}
 \label{tab:resultsDiffetas}
\end{table}

\begin{table}[h!]
 \centering
 \begin{tabular*}{1.\textwidth}{@{\extracolsep{\fill}}lcccc}
  \hline\hline
  ensemble & $aM_{K}^{K}$ & $aM_{K}^{\eta_s}$ & $aM^{K}_{\eta_s}$ & $aM_{\eta_s}^{\eta_s}$ \\ 
  \hline\hline
  $A40.24$   & $0.2583(16)$ & $0.2031(17)$ & $0.3824(13)$ & $0.3046(18)$  \\
  $A60.24$   & $0.2669(09)$ & $0.2161(09)$ & $0.3838(10)$ & $0.3123(14)$  \\
  $A80.24$   & $0.2770(06)$ & $0.2297(06)$ & $0.3860(08)$ & $0.3126(12)$  \\
  $A100.24$  & $0.2880(08)$ & $0.2422(09)$ & $0.3923(11)$ & $0.3178(16)$  \\
  \hline
  $A80.24s$  & $0.2551(07)$ & $0.2027(07)$ & $0.3533(09)$ & $0.2720(12)$  \\
  $A100.24s$ & $0.2649(07)$ & $0.2173(07)$ & $0.3538(12)$ & $0.2695(18)$  \\
  \hline
  $B55.32$   & $0.2280(04)$ & $0.1893(04)$ & $0.3221(04)$ & $0.2640(06)$  \\
  $D45.32sc$   & $0.1758(10)$ & $0.1618(10)$ & $0.2319(04)$ & $0.2096(05)$  \\
  \hline\hline
 \end{tabular*}
 \caption{OS kaon and $\eta_s$ mass values for both matching procedures.}
 \label{tab:OSmasses}
\end{table}

\begin{table}[h!]
 \centering
 \begin{tabular*}{1.\textwidth}{@{\extracolsep{\fill}}lcccccccccc}
  \hline\hline
  ensemble & $t_{1}^{\eta_\ell}$  & $t_{2}^{\eta_\ell}$ & $t_{1}^{\eta_s}$  & $t_{2}^{\eta_s}$ & $t_{0}^\eta$ &  $t_{1}^\eta$  & $t_{2}^\eta$ & $t_{0}^{\eta'}$ & $t_{1}^{\eta'}$  & $t_{2}^{\eta'}$  \\ 
  \hline\hline
  $A$-ensembles\hspace{-0.2cm} & 12 & 22 & 12 & 22 & 1 &  3 & 12 & 1 & 2 & 5 \\
  $B55.32$                     & 15 & 25 & 15 & 25 & 1 &  3 & 16 & 1 & 2 & 5 \\
  $D45.32sc$                   & 18 & 30 & 18 & 30 & 1 &  3 & 16 & 1 & 2 & 5 \\
  \hline
  $A80.24^{6\times 6}$   & & & & & 2 &  7 & 15  &  1 & 2 & 8  \\
  \hline\hline
 \end{tabular*}
 \caption{List of fit parameters. $t_{1}^{\eta_{l,s}}$ and
   $t_{2}^{\eta_{l,s}}$ define the fit intervals for the ground state
   of the connected correlation function in the light (strange)
   sector, which is required for the subtraction of excited states from
   the full correlator, before solving the GEVP for the resulting
   correlation function matrix. The $t_0$ values for the GEVP used to
   determine the $\eta$, $\eta'$ states are given by
   $t_0^{\eta,\eta'}$ whereas $t_1^{\eta,\eta'}$, $t_2^{\eta,\eta'}$
   denote the respective fit ranges to the principal correlators. In
   the last row we give the parameters for the GEVP applied to the
   full 6x6 correlation function matrix of the $A80.24$ ensemble for
   the $\eta_s$ matching case, as shown in the left panel of
   figure~\ref{fig:mA80ConnExcTrick}.} 
 \label{tab:fits}
\end{table}

\end{appendix}
\end{document}